\documentclass{jpp}
\usepackage{graphicx}
\usepackage{epstopdf, epsfig}

\newcommand{\be}{\begin{equation}}
\newcommand{\ee}{\end{equation}}
\newcommand{\ba}{\begin{eqnarray}}
\newcommand{\ea}{\end{eqnarray}}
\newcommand{\bas}{\begin{eqnarray*}}
\newcommand{\eas}{\end{eqnarray*}}

\newcommand{\vc}[1]{
{\bf{#1}}
}

\newcommand{\intl}{
\int \limits
}

\newcommand{\df}{
{\rm d}
}

\newcommand{\totd}[2]{
\frac{\df #1}{\df #2}
}

\newcommand{\eps}{\epsilon}

\newcommand{\kp}{\kappa}

\newcommand{\G}{\Gamma}

\shorttitle{Gyrokinetic Stability of Electron-Positron-Ion Plasmas}
\shortauthor{A.~Mishchenko, A.~Zocco, P.~Helander, and A.~K\"onies}

\title{Gyrokinetic Stability of Electron-Positron-Ion Plasmas}

\author{A.~Mishchenko\aff{1}
  \corresp{\email{alexey.mishchenko@ipp.mpg.de}}, 
  A.~Zocco\aff{1},
  P.~Helander\aff{1},  
  \and  A.~K\"onies\aff{1}}

\affiliation{\aff{1}Max Planck Institute for Plasma Physics,
  D-17491 Greifswald, Germany}

\begin{document}

\maketitle

\begin{abstract}
The gyrokinetic stability of electron-positron plasmas contaminated by ion (proton)
admixture is studied in slab geometry. 
The appropriate dispersion relation is derived and solved. 
The ion-temperature-gradient driven instability, 
the electron-temperature-gradient driven 
instability, the universal mode, and the shear Alfv\'en wave are
considered. The contaminated plasma remains stable if the contamination degree
is below some threshold, and it is found that the shear Alfv\'en wave can be present in a 
contaminated plasma in cases where it is absent without ion contamination.
\end{abstract}

\section{Introduction}

The prospects of creating electron-positron pair plasmas magnetically confined
in dipole or stellarator geometries have been discussed since early 2000's
\citep{pedersen2003}. In near future, the first experiment aiming at this goal will
be constructed \citep{pedersen2012}. Recently, efficient injection and
trapping of a cold positron beam in a dipole magnetic field configuration has
been demonstrated by \citet{saitoh2015}. This result is a key step towards the
ultimate aim of creating and studying of the first man-made
magnetically-confined pair plasma in the laboratory.  %dipole fields.  

It has been shown by \citet{per_prl} that pair plasmas possess unique
gyrokinetic stability properties thanks to the mass symmetry between the
particle species. For example, drift instabilities are completely absent
in straight unsheared geometry, e.~g.~in a slab. They can be destabilised only
in the presence of magnetic curvature in more complicated confining fields.  %Later,
\citet{per_jpp} found that this result persists also in the
electromagnetic regime. But, what 
happens if the perfect mass symmetry between the positively charged particles 
(positrons) and the negatively charged ones (electrons) is broken? This
can happen if some fraction of ions (e.~g.~protons) is introduced into the pair
plasma, which probably will be the case in experiments since the pumping and
vacuum systems are never completely perfect. 
Then one could expect that the drift instabilities will reappear. 

In this paper, we address the effect
of proton contamination on the gyrokinetic stability of pair plasmas. 
We find that drift instabilities can indeed appear in contaminated
pair plasmas if the proton fraction exceeds some threshold. Also, we find that
the shear Alfv\'en wave is present in contaminated plasma even if the ion
contamination is small. Its frequency, however, increases rapidly when the ion
fraction becomes negligible.  
%In pure pair plasma, the displacement current must be taken into 
%account when adressing the shear Alfv\'en wave \citep{Zocco_alf}. 

The structure of the paper is as follows. In \S\ref{Dispersion_relation}, the
general electromagnetic dispersion relation is derived. It describes slab
gyrokinetic stability in plasmas with an arbitrary number of species, although we
consider only three species in this work. In \S\ref{K-modes}, the stable
part of the gyrokinetic spectrum is addressed. 
In \S\ref{universal_mode}, \S\ref{ITG_mode} and \S\ref{ETG_mode}, drift
instabilities in three-component plasmas are considered. In
\S\ref{alfven_wave}, the shear Alfv\'en wave in electron-positron-ion
plasmas is described. Conclusions are summarised in \S\ref{sec:conclusions}. 
%
%================================================================= 
%
\section{Dispersion relation} \label{Dispersion_relation}
Following \citet{per_prl} and \citet{per_jpp}, we use gyrokinetic theory to
analyse the linear stability of electron-positron-ion plasmas. It is
convenient to write the gyrokinetic distribution function in the form:
\be
f_a = f_{a0} \left(1 - \frac{e_a \phi}{T_a}\right) + g_a = f_{a0} + f_{a1} \ , \;\;
f_{a1} = \,-\,\frac{e_a \phi}{T_a}\,f_{a0} + g_a 
\ee
Here, $f_{a0}$ is a Maxwellian, $a$ is the species index with $a=e$
corresponding to electrons, $a = p$ to positrons, and $a = i$ to
the ions. The linearised gyrokinetic equation in this notation is 
\be
i v_{\|} \nabla_{\|} g_a + (\omega - \omega_{{\rm d}a}) g_a  = \frac{e_a}{T_a}
\, J_0\left(\frac{k_{\perp}v_{\perp}}{\omega_{{\rm c}a}}\right) \, 
\Big(\omega - \omega^T_{*a}\Big) \, (\phi - v_{\|} A_{\|}) \, f_{a0}
\ee
with $J_0$ the Bessel function, $\omega_{{\rm c}a}$ the cyclotron frequency,
$k_{\perp}$ the perpendicular wave number, $\phi$ the perturbed electrostatic
potential and $A_{\|}$ the perturbed parallel magnetic potential in the
Coulomb gauge. Other notation used is
\ba
&&{} \omega_{*a}^T = \omega_{*a} \left[ 1 + 
\eta_a \left( \frac{v^2}{v_{{\rm th}a}^2} - \frac{3}{2}\right) \right] \ , \;\;
v = \sqrt{v_{\|}^2 + v_{\perp}^2} \ , \;\; k_{\perp} = \sqrt{k_x^2 + k_y^2} \\
&&{} \omega_{*a} = \frac{k_y T_a}{e_a} \totd{\ln n_a}{\psi} \ , \;\;
\eta_a = \totd{\ln T_a}{\ln n_a} \ , \;\; 
v_{{\rm th}a} = \sqrt{\frac{2 T_a}{m_a}} \ , \;\;
\omega_{{\rm d}a} = \vc{k}_{\perp} \cdot \vc{v}_{{\rm d}a} 
%\ , \;\;\vc{v}_d = \frac{m v_{\|}^2}{q B} (\nabla \times \vc{b})_{\perp} 
\ea
Here, the sign convention is such that $\omega_{*i} \le 0$, 
$\omega_{*p} \le 0$, and $\omega_{*e} \ge 0$.
For simplicity we will assume $k_x = 0$ and $k_{\perp} = k_y$ throughout the paper. 
In slab geometry, $\omega_{{\rm d}a} = 0$. Taking the Fourier transform along the
parallel coordinate, we obtain:
\be
( \omega - k_{\|} v_{\|} ) g_a  = \frac{e_a}{T_a}
\, J_0\left(\frac{k_{\perp}v_{\perp}}{\omega_{{\rm c}a}}\right) \, 
\Big(\omega - \omega^T_{*a}\Big) \, (\phi - v_{\|} A_{\|}) \, f_{a0}
\ee
This equation is trivially solved: 
\be
g_a = \frac{\omega - \omega^T_{*a}}{\omega - k_{\|} v_{\|}} \frac{e_a
  f_{a0}}{T_a} \, J_0 \, (\phi - v_{\|}A_{\|}) 
\ee
The gyrokinetic quasineutrality condition and the parallel Ampere's law are 
\be
\left( \sum_a \frac{n_a e_a^2}{T_a} + \eps_0 \, k_{\perp}^2 \right) \phi
= \sum_a e_a \int g_a J_0 \df^3 v \ , \;\;
A_{\|} = \frac{\mu_0}{k_{\perp}^2} \sum_a e_a \int v_{\|} g_a J_0 \df^3 v
\ee

For the electromagnetic dispersion relation, it is convenient to define: 
\be
\label{I_na_def1}
W_{n a} = \,-\, \frac{1}{n_a v_{{\rm th}a}^n} \int \frac{\omega - \omega^T_{*a}}{\omega - k_{\|} v_{\|}} \,
J_0^2 \, f_{a0} \, v_{\|}^n \, \df^3 v 
\ee
Taking velocity-space integrals, one finds:
\be
\label{I_na_def2}
W_{na} = \zeta_a \left\{ \left( 1 - \frac{\omega_{*a}}{\omega} \right) Z_{na} \G_{0a} +
  \frac{\omega_{*a} \eta_a}{\omega} \, \left[ \frac{3}{2} Z_{na} \G_{0a} - 
Z_{na} \G_{*a} - Z_{n+2,a} \G_{0a} \right]\right\} 
\ee
Here, the following notation is employed:
\ba
&&{} \frac{1}{\lambda_{{\rm D}a}^2} = \frac{q_a^2 n_a}{\eps_0 T_a} \ , \;\; 
\frac{1}{\lambda_{\rm D}^2} = \sum_a \frac{1}{\lambda_{{\rm D}a}^2} \ , \;
b_a = k_{\perp}^2 \rho_a^2 \ , \; \rho_a = \frac{\sqrt{m_a T_a}}{|q_a| B} \\
&&{} \G_{*a} = \G_{0a} - b_a \Big[\G_{0a} - \G_{1a}\Big]  \ , \;\; 
\G_{0a} = I_0(b_a) e^{-b_a} \ , \;\; \G_{1a} = I_1(b_a) e^{-b_a} \\
&&{} Z_{na} = \frac{1}{\sqrt{\pi}} \intl_{-\infty}^{\infty} \frac{x^n e^{-x^2}
  \df x}{x - \zeta_a} \ , \;\; 
\zeta_a = \frac{\omega}{k_{\|} v_{{\rm th}a}} 
\ea
Using this notation, we can cast the field equations into the form: 
\ba
&&{} \Big(1 + k_{\perp} \lambda_D^2\Big) \phi + \sum_a \frac{\lambda_D^2}{\lambda_{Da}^2} \Big(
W_{0a} \, \phi  - W_{1a} \, A_{\|} v_{{\rm th}a} \Big) = 0 \\
&&{} A_{\|} + \frac{1}{c^2} \sum_a \frac{v_{{\rm th}a}}{k_{\perp}^2 \lambda_{Da}^2} \Big(
W_{1a} \, \phi  - W_{2a} \, A_{\|} v_{{\rm th}a} \Big) = 0
\ea
Computing the determinant of this system of equations, we find the
electromagnetic dispersion relation describing electron-positron-ion plasma in
slab geometry:
\ba
\label{disp_rel}
&&{} \left( 1 + k_{\perp}^2 \lambda_D^2 +
\sum_a \frac{\lambda_D^2}{\lambda_{Da}^2} \, W_{0a} \right) 
\left( 1 - 2 \sum_a \frac{\beta_a}{k_{\perp}^2 \rho_a^2} \, W_{2a} \right)
\,+\, \\
&&{} \,+\, 2 \sum_a \frac{\lambda_D^2}{\lambda_{Da}^2} \; W_{1a} v_{{\rm th}a} \; 
\sum_a \frac{\beta_a}{k_{\perp}^2 \rho_a^2} \; \frac{W_{1a}}{v_{{\rm th}a}} = 0 \nonumber
\ea
Here, $\beta_a = \mu_0 n_a T_a / B^2$. The electrostatic limit corresponds, as
usual, to $\beta_a = 0$. 

In the following, we will use this dispersion relation in order to describe
instabilities which can appear in three-component plasmas. This will give
us insight into the general properties of the gyrokinetic stability of such plasmas.  
%
%=================================================================
%
%
\newcommand{\gsize}{
7.0
}
\newcommand{\graph}[1]{
\begin{minipage}{\gsize cm}
%\vspace*{-0.4cm}
\hspace*{-1.0cm}
\epsfclipon
\epsfysize \gsize truecm
%\hfill 2 \hrulefill 
\psfig{figure=../graph/DISPERS/slab/#1.eps,width=\gsize cm}
%\hfill
\end{minipage}
}
\section{Gyrokinetic stable modes} \label{K-modes}
\begin{figure}
  \vspace{1.0cm}
    \centerline{
      \includegraphics[width=7cm]{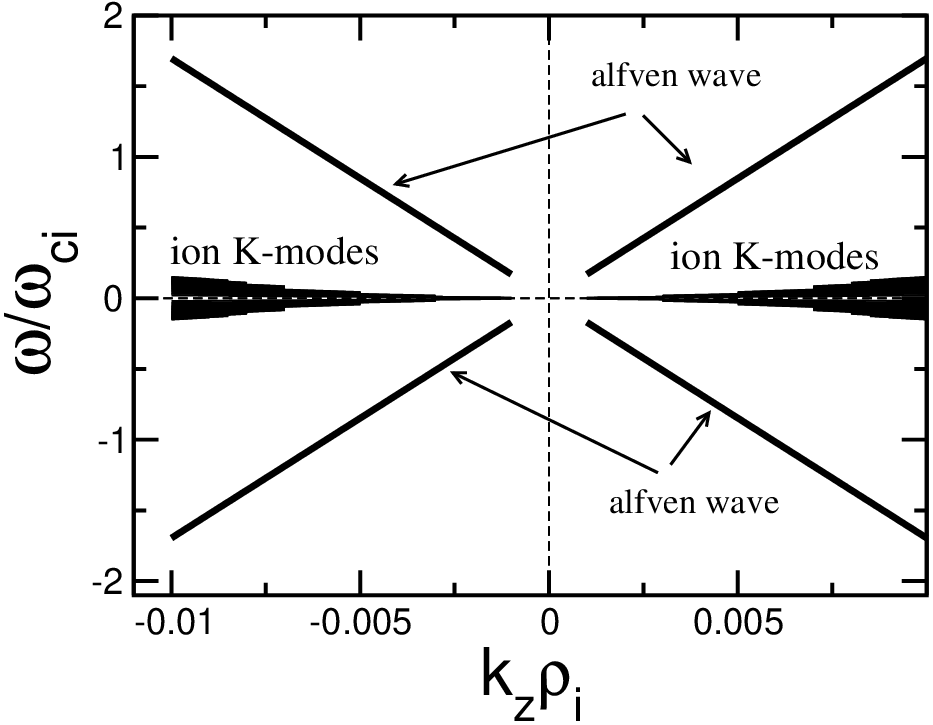}
%\vspace{0.5cm} 
\hspace{1.0cm}
      \includegraphics[width=7cm]{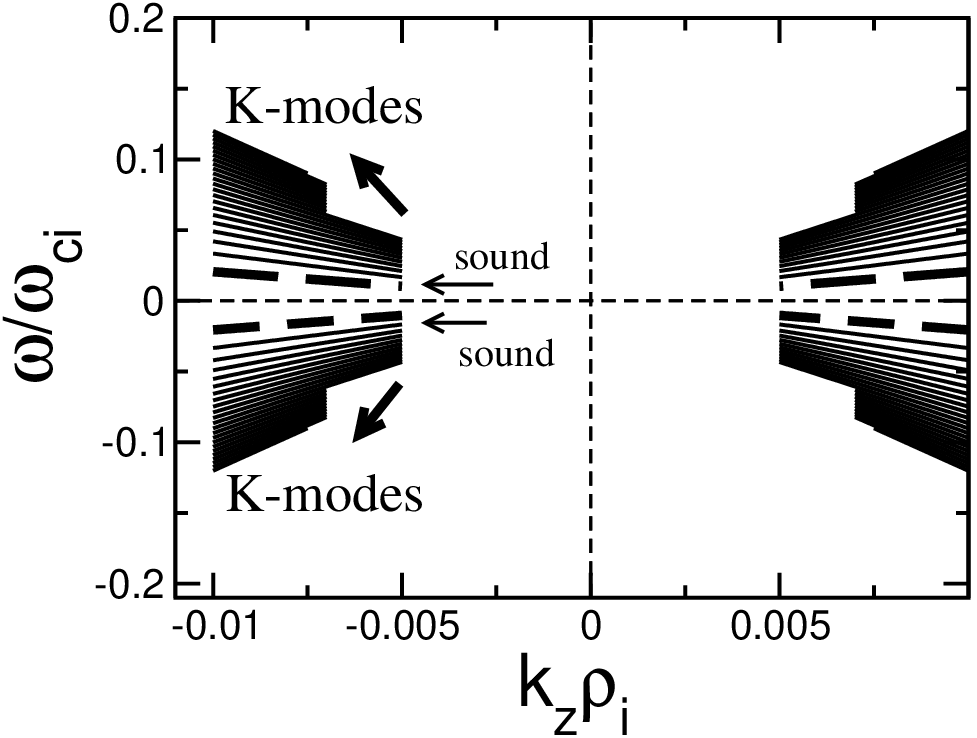} }
\caption{Left: gyrokinetic frequency spectrum for conventional plasmas
  including sound and Alfv\'en waves. Right: low-frequency part of the spectrum.} 
\label{fig:kmodes}
\end{figure}

\begin{figure}
  \vspace{1.0cm}
    \centerline{
      \includegraphics[width=7cm]{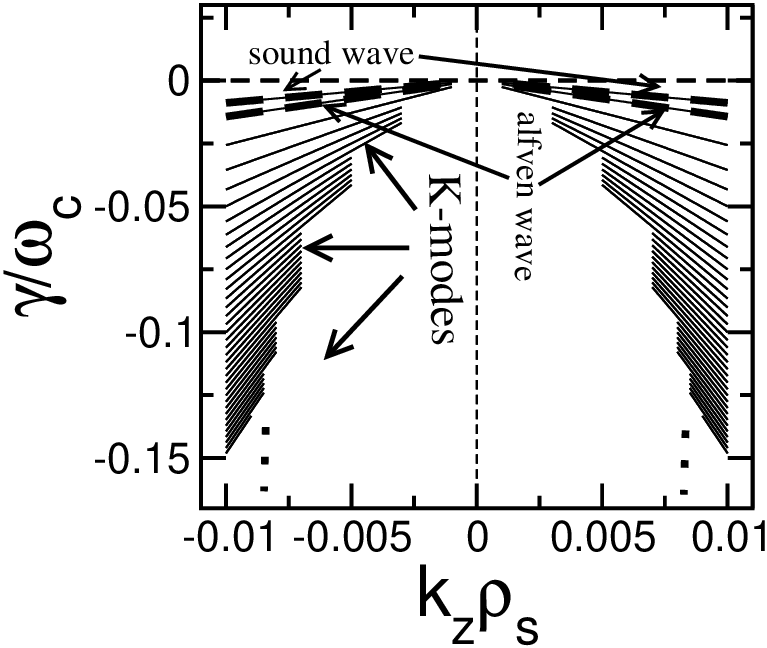}
\hspace{1.0cm}
      \includegraphics[width=7cm]{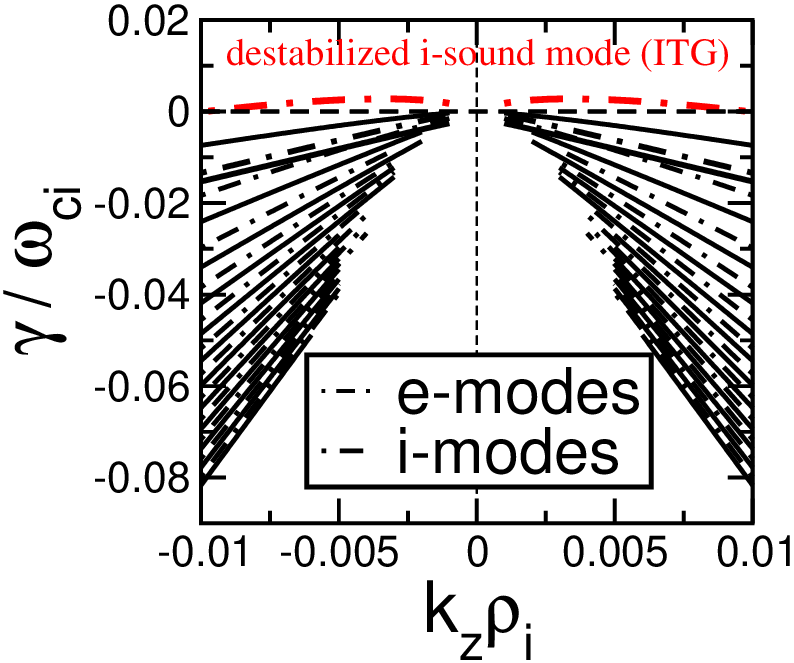} }
\caption{Left: the imaginary part of the spectrum in a homogeneous
plasma. Right: the same in the presence of an ion temperature gradient
$\kp_{{\rm T}i} = 0.1$. In this Figure, i-modes denote modes rotating in the
ion diamagnetic direction and e-modes correspond to modes rotating in the
electron diamagnetic direction.} 
\label{diam}
\end{figure}

We first consider the case of a pure electrostatic electron-positron
plasma. Assuming quasineutrality $\omega_{*p} = \,-\,\omega_{*e}$, 
equal temperatures $T_p = T_e$, and equal temperature gradients $\eta_p =
\eta_e$, we can reduce the dispersion relation to 
\be
\label{Kmode_disp}
1 + k_{\perp}^2 \lambda_{\rm D}^2 + \zeta Z_0 = 0 
\ee
Equations of this type have been considered in detail by \citep{Fried_Kmodes,UkrSSR,JETP}
for conventional (hydrogen) plasmas. In a hydrogen plasma,
equation (\ref{Kmode_disp_H}), similar to Eq.~(\ref{Kmode_disp}), describes
the plasma stability in the
absence of the density and temperature gradients and assuming $T_i = T_e$:
\be
\label{Kmode_disp_H}
1 + k_{\perp}^2 \lambda_{\rm D}^2 + \frac{1}{2}\left[\zeta_i Z_0(\zeta_i) \G_{0i} +
  \zeta_e Z_0(\zeta_e) \G_{0e} \right] = 0 
\ee
This equation has an infinite number of solutions, called {\it 
  K-modes} \citep{UkrSSR,JETP}. These modes can be of the ion type with $\zeta_i \ge 1$ or the
electron type with $\zeta_e \ge 1$. 
In Fig.~\ref{fig:kmodes}, the spectrum resulting from
Eq.~(\ref{Kmode_disp_H}) for the conventional plasma is plotted including 
K-modes of the ion type. This spectrum was computed numerically using the
Nyquist technique \citep{nyquist1,nyquist2}. The staircase-like visual appearance
of Figs.~\ref{fig:kmodes} and \ref{diam} is an artefact caused by the density
of the roots of the dispersion relation increasing towards the origin of the
coordinates. This complicates the numerical solution in this area. 

In Fig.~\ref{diam}, one sees that, as \citet{Fried_Kmodes} suggested, most of the
solutions of Eq.~(\ref{Kmode_disp_H}) are strongly damped, satisfying 
$|\gamma| \sim |\omega|$. The least damped solutions can be
destabilised by plasma profile gradients leading either to the
Ion Temperature Gradient driven instability (ITG), or the Electron Temperature
Gradient driven instability (ETG), or the universal instability, driven
by the density gradient. This is shown in Fig.~\ref{diam}, where the effect of
the ion temperature gradient on the gyrokinetic spectrum in a conventional
plasma can be seen. 
In pure pair plasmas, however, the electron and the positron
diamagnetic contributions cancel also in presence of profile gradients, 
making such plasmas absolutely stable in slab 
geometry within the gyrokinetic description. Note however, that perfect
symmetry between the electron and positron density and temperature profiles is
required to guarantee the cancellation of the diamagnetic terms. While density
profiles are always identical for the two species in a quasineutral plasma, 
the temperature profiles can differ. In this case, a pure pair plasma can be
temperature-gradient unstable, as we will see in the following. The gradient-driven
instabilities can also appear if a pair plasma is ``contaminated'' by protons
or other ions. 

\begin{figure}
  \vspace{1.0cm}
    \centerline{
      \includegraphics[width=7cm]{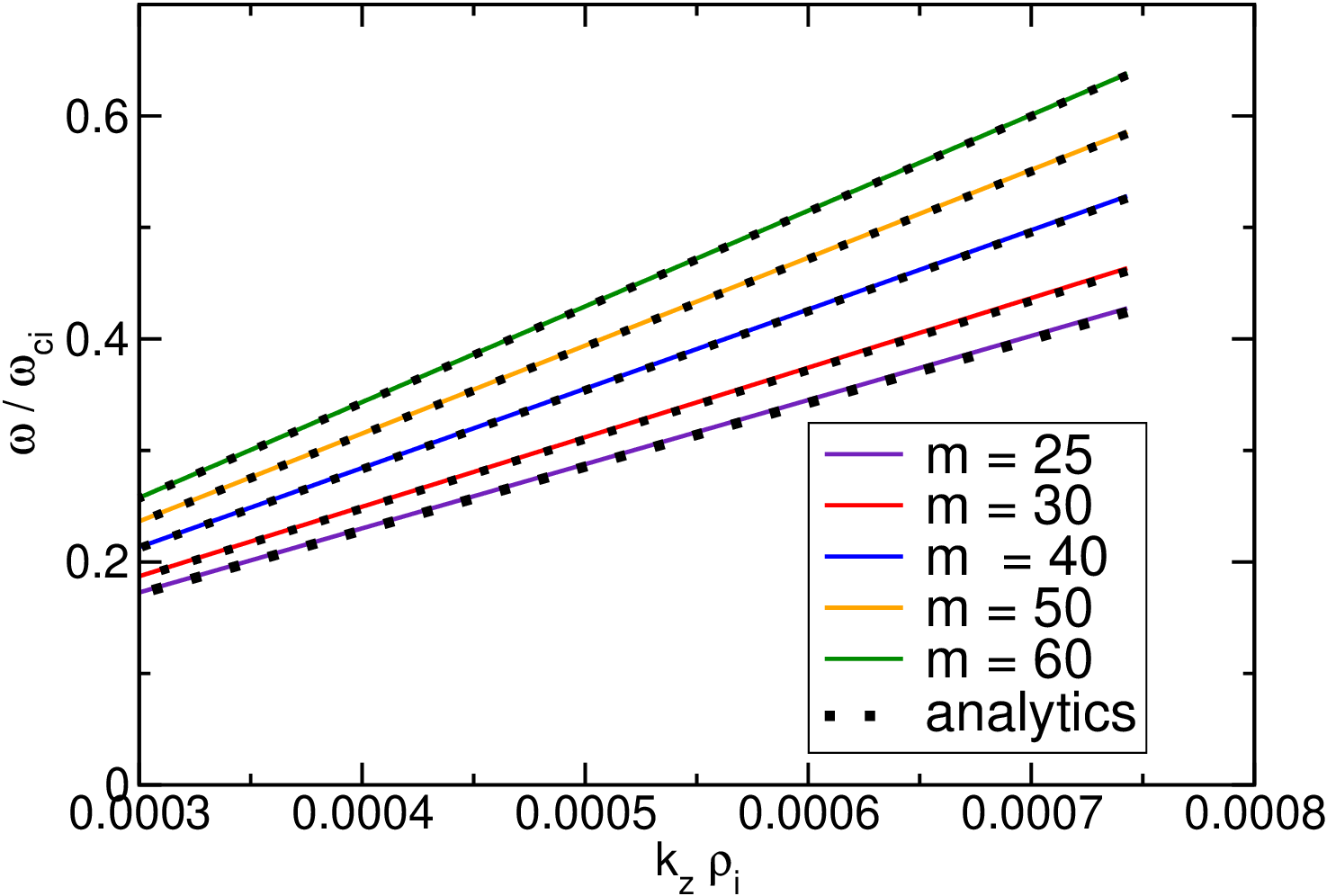}
%\vspace{0.5cm} 
\hspace{1.0cm}
      \includegraphics[width=7cm]{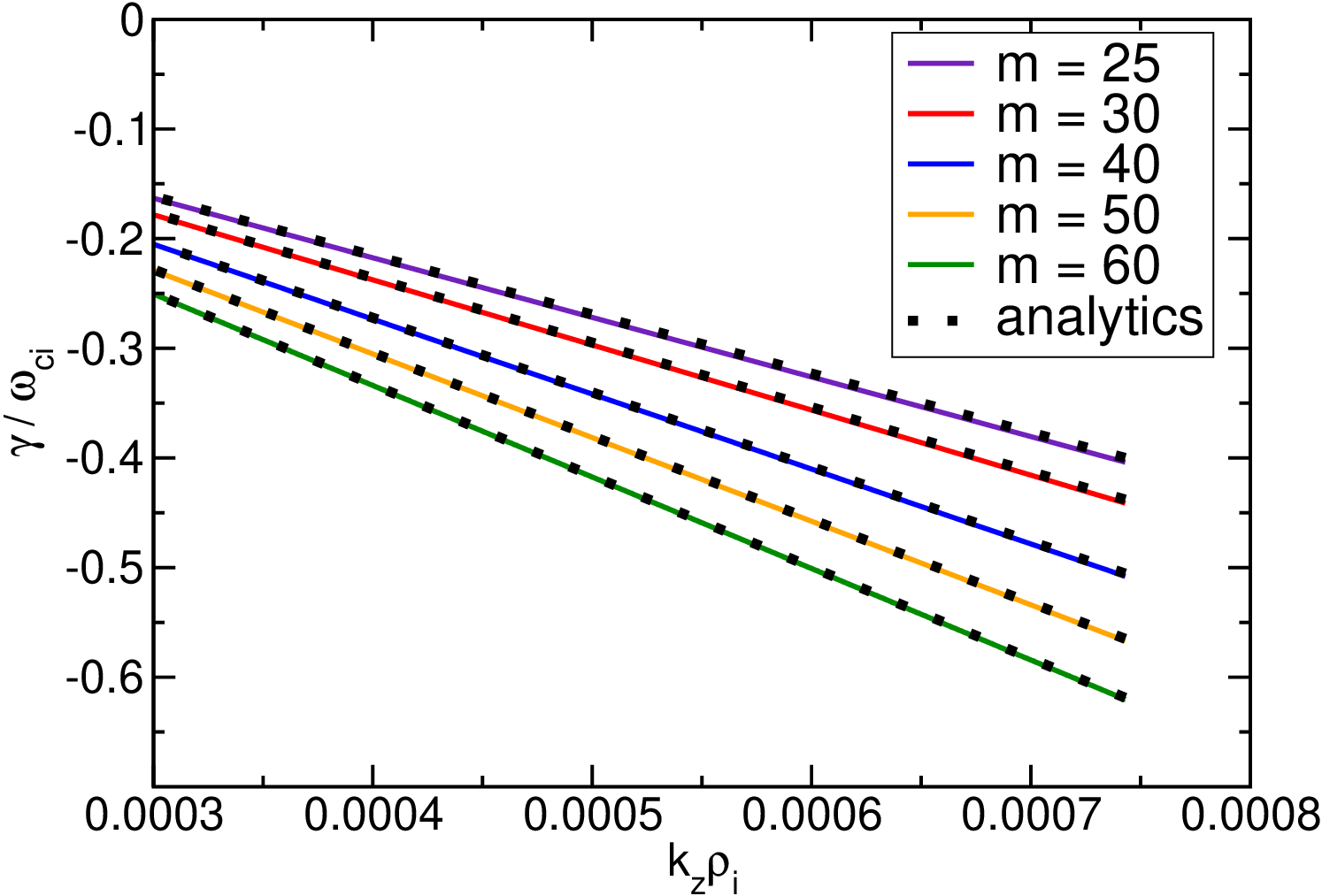} }
\caption{``K-mode'' solution of the dispersion relation for a pure pair
   plasma. All modes are strongly damped. Here, $k_{\perp} \lambda_D = 0$ has
   been assumed. The numerical solution of Eq.~(\ref{disp_rel}) is compared with the analytic
   result Eq.~(\ref{Kmodes_pair_sol}).}
\label{kmodes_analyt}
\end{figure}

Some analytic progress can be made for K-modes in an electron-positron plasma. 
Assuming $\zeta_e = \zeta_p \gg 1$ and $\gamma \sim \omega$, the plasma dispersion
function can be approximated:
\be
\label{kmode_exand}
Z_0(\zeta_e) \approx 2 i \sqrt{\pi} e^{-\zeta_e^2} - \frac{1}{\zeta_e} - \frac{1}{2 \zeta_e^3}
\ee
For this expansion, the dispersion relation takes the form:
\be
\label{Kmodes_pair_disp}
4 i \sqrt{\pi} \zeta_e^3 e^{-\zeta_e^2} = 1
\ee
Introducing the notation $\zeta_e = x - i y$ and assuming $x = \pm (y + \Delta)$ with
$\Delta \ll y$ \citep{UkrSSR,JETP}, we can write the dispersion relation in the form:
\be
8 y^3 \sqrt{2 \pi} e^{-2 y \Delta} \; \exp(2 i y^2 - i \pi/4) = 1 \equiv
\exp(2 \pi m i)
\ee
Splitting this relation into equations for the argument and for the
absolute value and employing $\Delta/y \ll 1$, we obtain:
\be
2 y^2 - \pi/4 = 2 \pi m \ , \;\; 8 y^3 \sqrt{2 \pi} e^{-2 y \Delta} = 1
\ee
Thus, an infinite family of solutions is found:
\be
y_m = \sqrt{\pi m + \pi/8} \approx \sqrt{\pi m} \ , \;\;
\Delta_m = \frac{\ln(8 y_m^3 \sqrt{2 \pi}) }{2 y_m} \ , \;\;
x_m = \pm (y_m + \Delta_m)
\ee
Finally, we write our solutions in the form:
\be
\label{Kmodes_pair_sol}
\omega_m = \pm k_{\|} v_{{\rm th}e} x_m   \ , \;\; 
\gamma_m = \,-\, k_{\|} v_{{\rm th}e} y_m 
\ee
These relations describe strongly-damped K-modes in a pure
electron-positron plasma. In Fig.~\ref{kmodes_analyt}, we compare these
analytic results with the numerical solution of the original dispersion
relation Eq.~(\ref{disp_rel}) and find very good agreement. 
Note that the expansion Eq.~(\ref{kmode_exand}) is valid for $m \gg 1$. 
For low $m$, the dispersion relation must be solved numerically. 
%using, for instance, the Nyquist technique. 

\begin{figure}
\vspace{1.0cm}
    \centerline{
      \includegraphics[width=7cm]{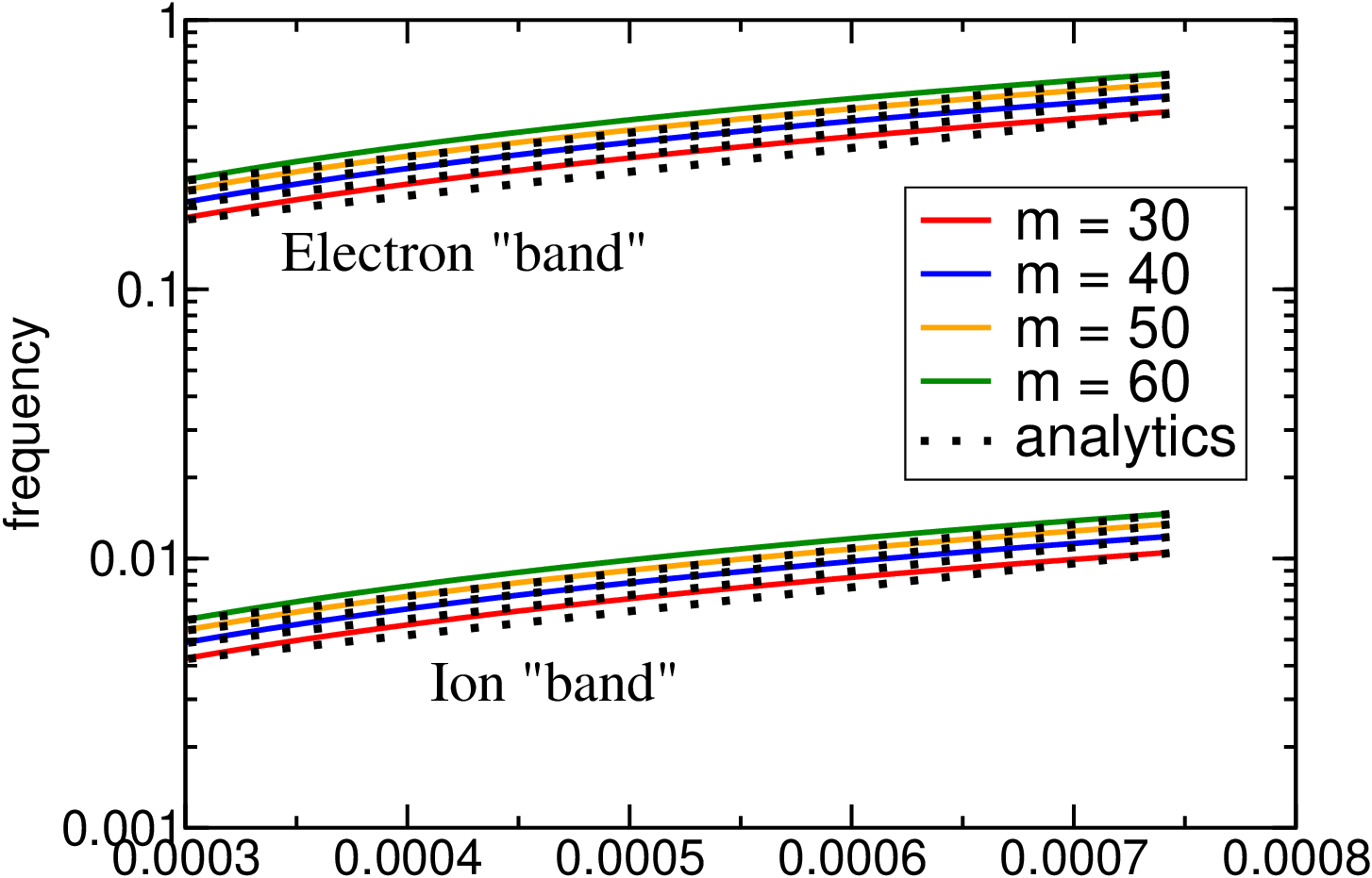}
%\vspace{0.5cm} 
\hspace{1.0cm}
      \includegraphics[width=7cm]{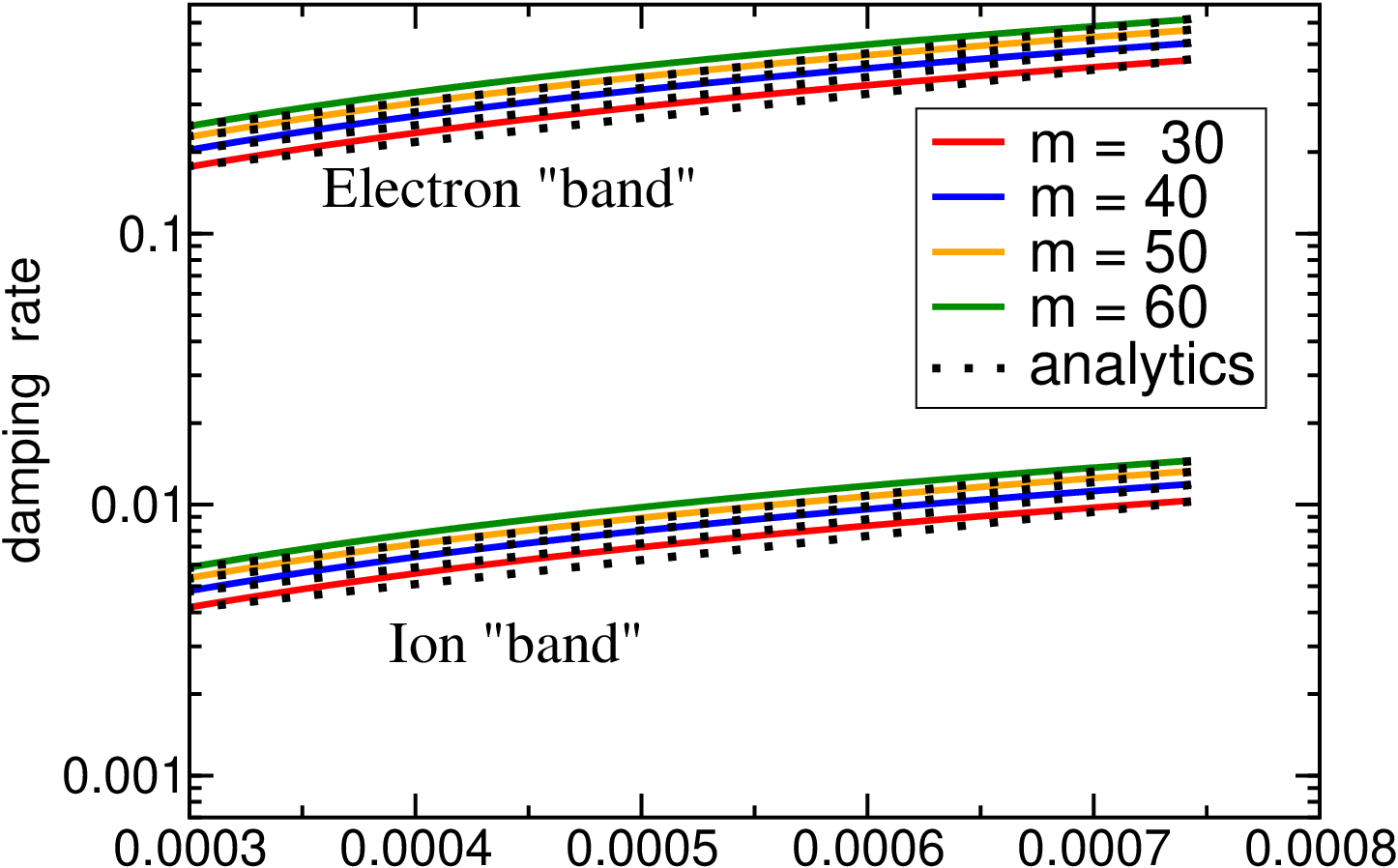} }
\caption{``K-mode'' solution of the dispersion relation for conventional
  plasma assuming $k_{\perp} \lambda_D = 0$. One can see the ion and the
  electron parts of the spectrum. The numerical solution of
  Eq.~(\ref{disp_rel}) is compared with the analytic result
  Eq.~(\ref{Kmodes_H_sol}).} 
\label{kmodes_H_analyt}
\end{figure}

In a conventional plasma, one can make the usual assumption $\zeta_i \gg 1$ and
$\zeta_e \ll 1$. In this case, the following expansions can be used:
\be
Z_0(\zeta_i) = 2 i \sqrt{\pi} e^{-\zeta_i^2} - \frac{1}{\zeta_i} - 
\frac{1}{2 \zeta_i^3} \ , \;\;
Z_0(\zeta_e) = i \sqrt{\pi} - 2 \zeta_e
\ee
which lead to the approximated dispersion relation:
\be
\left(1 - \G_{0i}/2 \right) + \left(i \zeta_i \sqrt{\pi} e^{-\zeta_i^2} -
  \frac{1}{4 \zeta_i^2}\right) \G_{0i} + {\cal O}(\zeta_e) = 0
\ee
For simplicity, we neglect Finite Larmor Radius (FLR) effects, implying
$\G_{0i} = 1$. Also, the small contribution $1/(4 \zeta_i^2) \ll 1$ can be
neglected compared to the other terms. Then, we obtain:
\be
\label{Kmodes_ion_disp}
2 i \zeta_i \sqrt{\pi} e^{-\zeta_i^2} + 1 = 0
\ee
Using the notation $\zeta_i = x - i y$ with $x = \pm (y + \Delta)$ and
employing $\Delta \ll 1$, we can split the dispersion relation into equations for 
the argument and for the absolute value:
\be 
2 y \sqrt{2 \pi} e^{-2 y \Delta} \exp(2 i y^2 - 3 \pi i / 4 ) = 1 \equiv
\exp(2 \pi m i)
\ee
Finally, the solutions for the K-modes of the ion type are 
\be
\label{Kmodes_H_sol}
y_m = \sqrt{\pi m + \frac{3 \pi}{8}} \approx \sqrt{\pi m} \ , \;\; 
\Delta_m = \frac{\ln(2 y \sqrt{2 \pi})}{2 y} \ , \;\;
x_m = y_m + \Delta_m
\ee
In Fig.~\ref{kmodes_H_analyt}, these analytic results are compared with the
numerical solution of the original (exact) dispersion relation Eq.(\ref{disp_rel}).  

Interestingly, the same dispersion relation can be obtained for K-modes in a pure pair plasma 
keeping the Debye length finite. In this case, the dispersion relation Eq.~(\ref{Kmodes_pair_disp}) is replaced by 
\be
\label{Kmodes_pair_disp_LD}
4 i \sqrt{\pi} \zeta_e^3 e^{-\zeta_e^2} + 2 \zeta_e^2 k_{\perp}^2 \lambda_D^2 = 1 
\;\; \Longrightarrow \;\;
2 i \sqrt{\pi} \zeta_e e^{-\zeta_e^2} + k_{\perp}^2 \lambda_D^2 = 0
\ee
which reduces to Eq.~(\ref{Kmodes_ion_disp}) if $k_{\perp} \lambda_D \gg 1 / \zeta_e$ 
with $k_{\perp} \lambda_D$ replacing $1$ and $\zeta_e$ replacing $\zeta_i$. 

In a hydrogen plasma, a dispersion relation, very similar to Eq.~(\ref{Kmodes_ion_disp}), can be obtained assuming $\zeta_e \gg 1$:
\be
\frac{1 - \G_{0i}}{2} + \left(i \zeta_e \sqrt{\pi} e^{-\zeta_e^2} -
  \frac{1}{4 \zeta_e^2}\right) + {\cal O}\left(\frac{1}{\zeta_i^2}\right) = 0
\ee
Here, recall that $\zeta_i \gg \zeta_e$. This dispersion relation coincides
with the ion K-mode dispersion relation Eq.~(\ref{Kmodes_ion_disp}) at finite
$k_{\perp} \rho_i$, and transforms into the pair-plasma K-mode dispersion
relation without $k_{\perp} \lambda_D$, see Eq.~(\ref{Kmodes_pair_disp}), when
$k_{\perp} \rightarrow 0$.  

In a three-component plasma with the ion fraction $\nu_i = n_i/n_e$, the
K-mode dispersion relation for $\zeta_e \gg 1$ becomes 
\be
\nu_i (1 - \G_{0i}) + (2 - \nu_i) \left(2 i \zeta_e \sqrt{\pi} e^{-\zeta_e^2} -
  \frac{1}{2 \zeta_e^2}\right) + {\cal O}\left(\frac{1}{\zeta_i^2}\right) = 0
\ee
The last term ($\sim 1/\zeta_e^2$) is negligible unless $\nu_i \rightarrow 0$ or
$k_{\perp} \rightarrow 0$. Here, electron and positron FLR effects have
been neglected.

In summary, K-modes, considered in this Section, are the only solutions of the
slab dispersion relation in pure 
electron-positron plasma for arbitrary density and temperature profiles
provided these profiles coincide for the two species. If the positron and
the electron temperature profiles
differ, a temperature-driven instability can appear also for pure pair plasma
in slab geometry. This will be considered in more detail in the following. 
%
%=================================================================
%
\section{Universal instability} \label{universal_mode}
\begin{figure}
  \vspace{1.0cm}
    \centerline{
      \includegraphics[width=7cm]{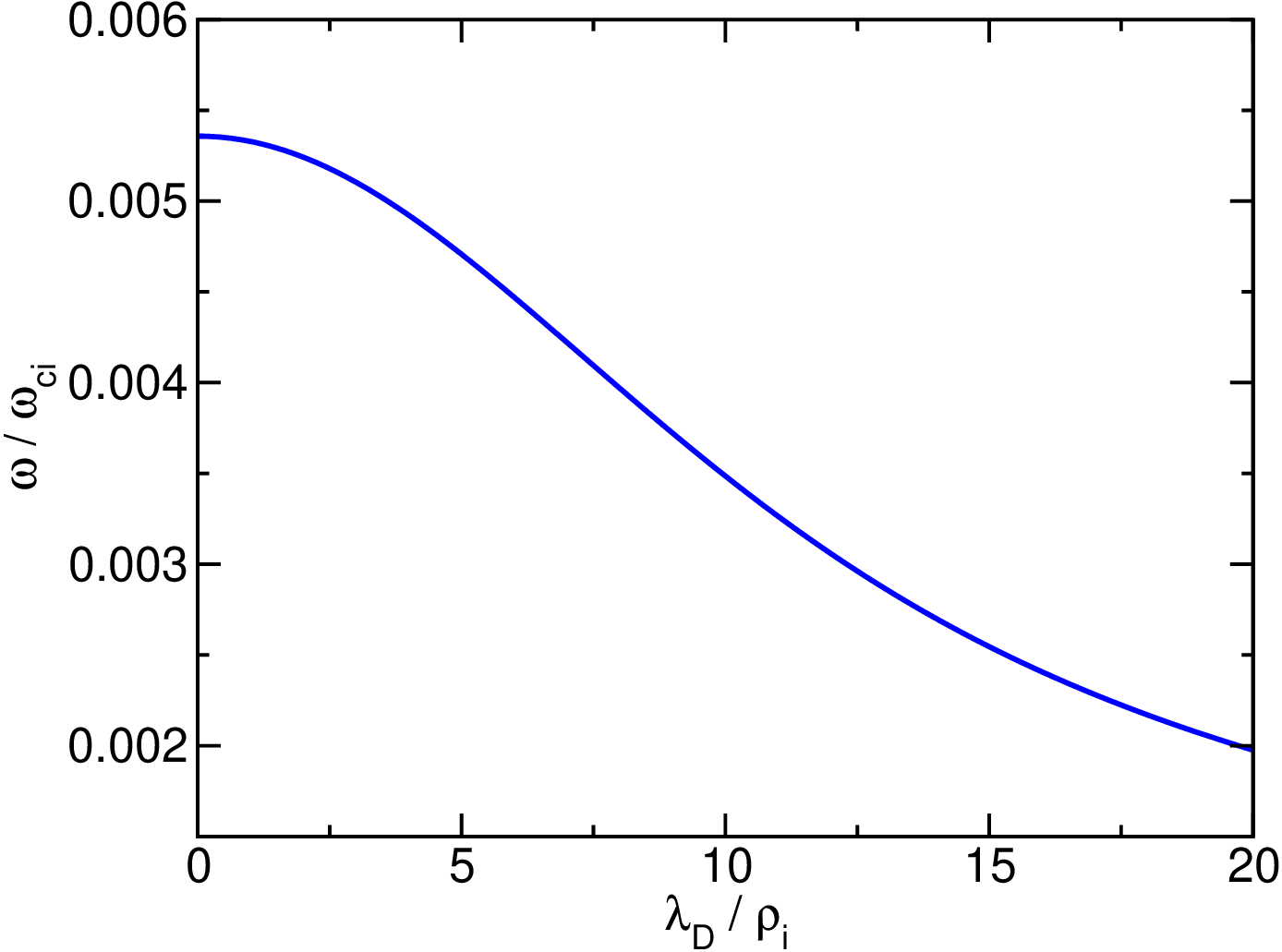}
      \hspace{0.2cm}
      \includegraphics[width=7cm]{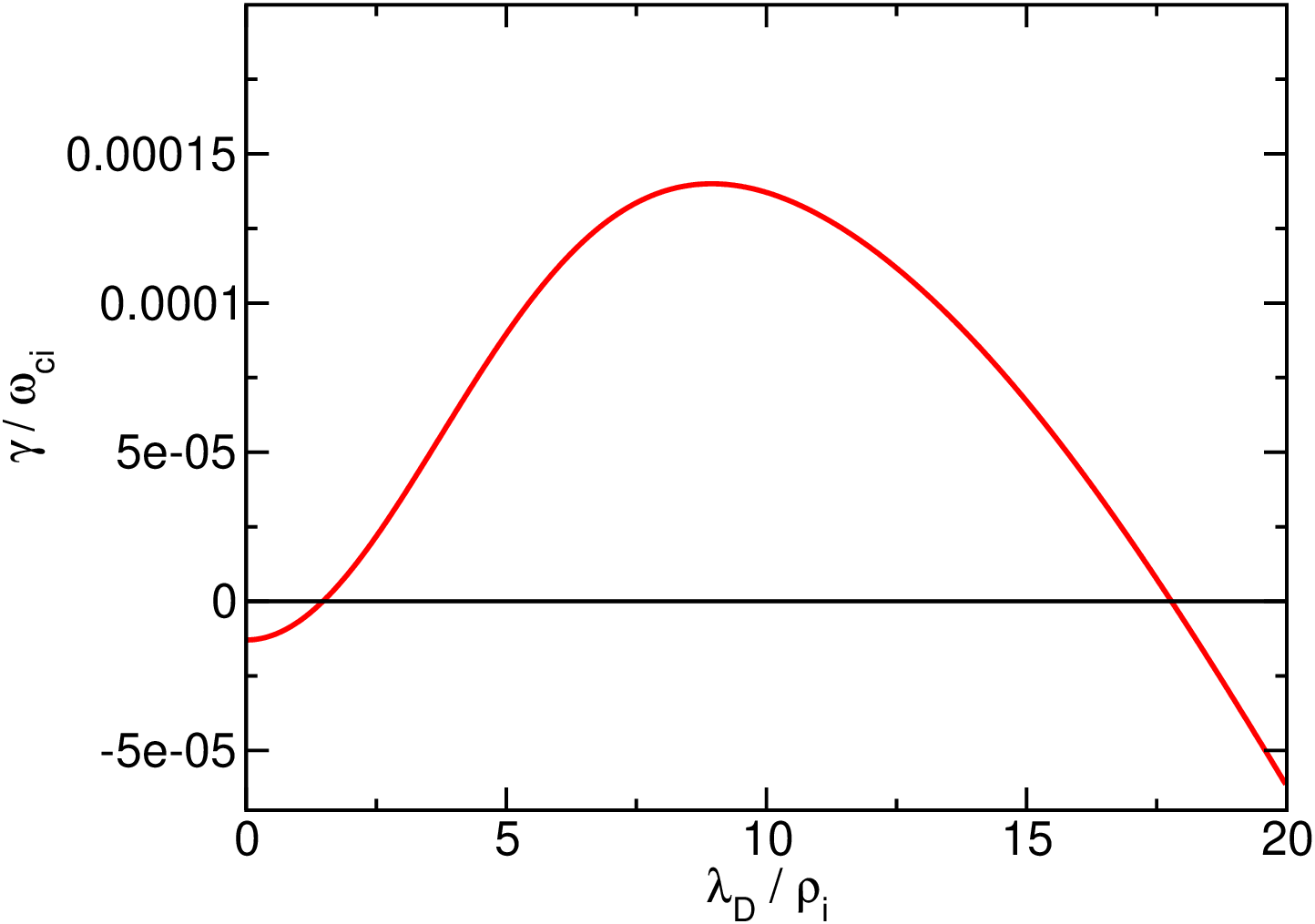} 
    }
\caption{Frequency and growth rate of the universal mode as functions 
  of the Debye length in a contaminated pair plasma with the positron fraction
  $\nu_p = 0.7$. 
%The numerical solution of Eq.~(\ref{disp_rel}) is compared with the analytic result Eq.~(\ref{w_univ}). 
The parameters are $k_{\perp} \rho_i = 0.1$, $k_{\|} \rho_i = 7.43 \times 10^{-4}$, 
$\kappa_{ni} \rho_i = \kappa_{ne} \rho_i = \kappa_{np} \rho_i = 0.3$, and $\kappa_{Ti} \rho_i = \kappa_{Te} \rho_i = \kappa_{Tp} \rho_i = 0.0$ with 
$\kappa_{na} = \df \ln n_a / \df \ln x$ and $\kappa_{Ta} = \df \ln T_a / \df \ln x$.}
\label{univ_analyt}
\end{figure}
\begin{figure}
  \vspace{1.0cm}
    \centerline{
      \includegraphics[width=7cm]{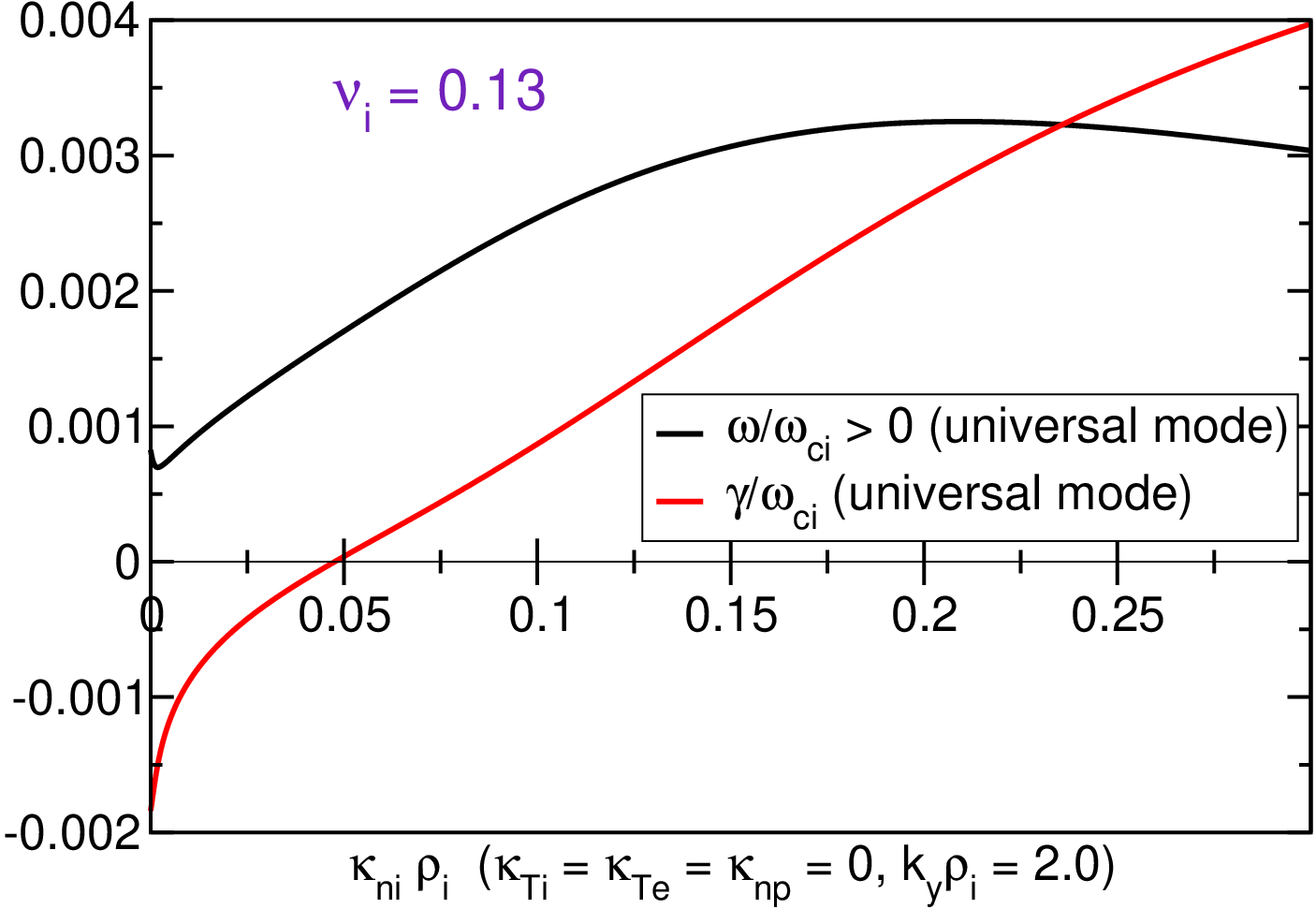}
      \hspace{0.2cm}
      \includegraphics[width=7cm]{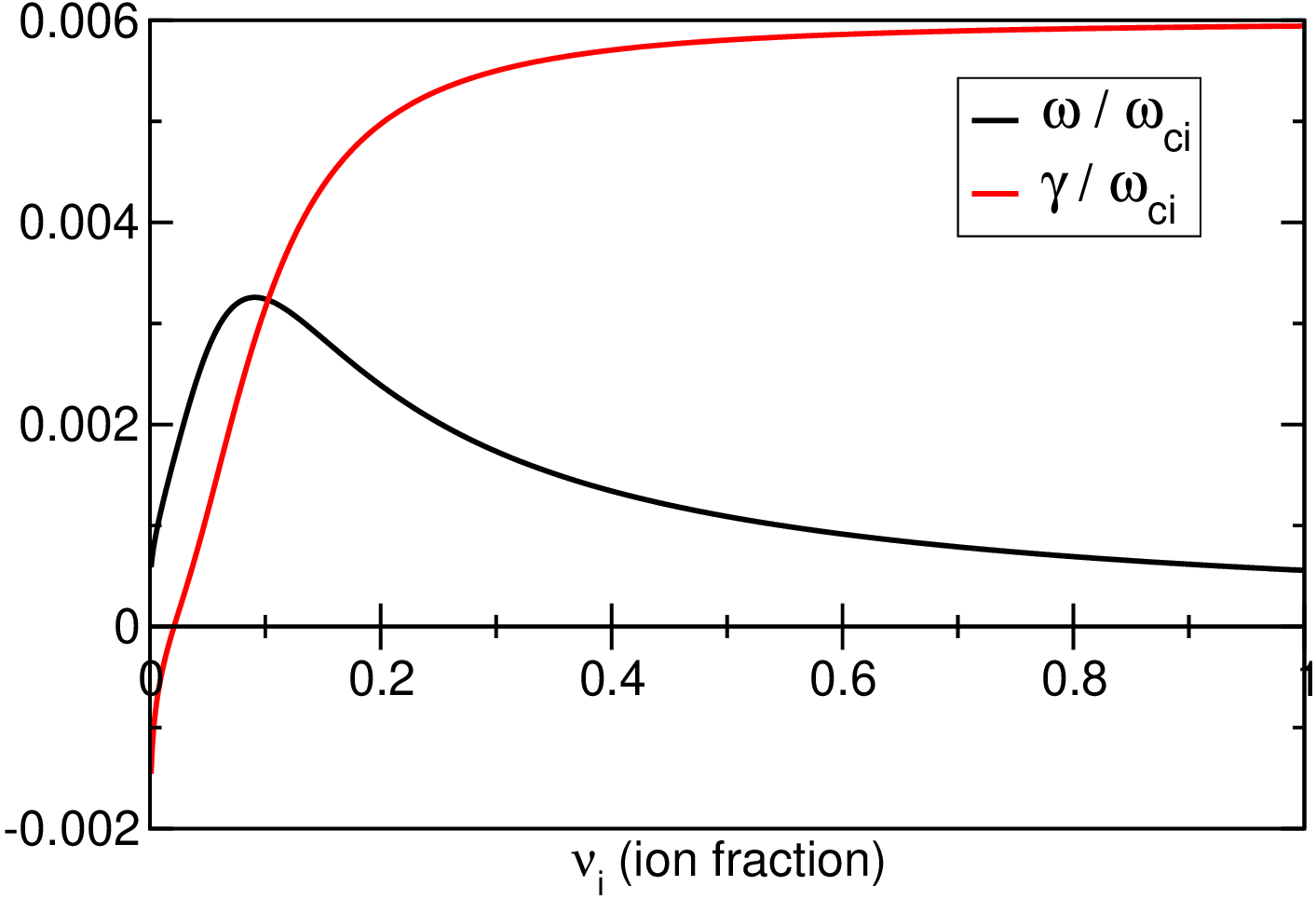} 
    }
\caption{Frequency and growth rate of the universal mode in a
  contaminated pair plasma. One sees that the ion density gradient and the ion
  contamination must be larger than some threshold for the mode to become
  unstable. The ion density gradient $\kappa_{ni} \rho_i = 0.3$ has been used
  for the $\nu_i$ dependence (figure on the right).}
\label{univ}
\end{figure}
The first unstable mode to be considered is the universal
instability driven by the density gradient. For simplicity, we assume
the temperature profiles to be flat. In this case, the dispersion relation is 
\be
1 + k_{\perp}^2 \lambda_D^2 + \frac{1}{2} \sum_{a=i,p,e} \nu_a \zeta_a \left(1 - \frac{\omega_{*a}}{\omega}\right)
Z_{0s} \G_{0s}= 0
\ee
Here, $\nu_a = n_a / n_e$ is the density fraction corresponding to a particular species $a = i,e,p$. For electrons, $\nu_e = 1$. 
Taking the limit $k_{\|} v_{{\rm th}i} \ll \omega \ll k_{\|} v_{{\rm th}e}$, we obtain:
\be
Z_{0i} \approx \,-\,\frac{1}{\zeta_i} \ , \;\; Z_{1e} \approx i \sqrt{\pi} 
\ee
To lowest order, the dispersion relation reduces to 
\be
\left[ 2 \left(1 + k_{\perp}^2 \lambda_D^2\right) - \nu_i \G_{0i} \right] \, \omega  - \nu_i \omega_* \G_{0i} +
i \zeta_e \sqrt{\pi} \Big[ \omega (\nu_e + \nu_p) - \nu_i \omega_*)\Big] = 0
\ee
Here, the notation $\omega_* = \omega_{*e} = \,-\,\omega_{*p}$ and 
quasineutrality, $\nu_e - \nu_p = \nu_i$,  have been used. 
The electron and positron FLR have been neglected $\G_{0e} = \G_{0p} = 1$. We solve the dispersion relation for $\omega = \omega_r + i\,\gamma$ 
assuming $\omega_r \gg \gamma$. Then, to the lowest order,
\be
\label{w_univ}
\omega_r = \frac{\nu_i \omega_* \G_{0i}}{2 \left(1 + k_{\perp}^2 \lambda_D^2\right) - \nu_i \G_{0i}}  \ , \;\; 
\gamma = 2 \zeta_e \sqrt{\pi} \, \nu_i \omega_*  \, \frac{k_{\perp}^2 \lambda_D^2 + (1 - \G_{0i})}{\left[2 \left(1 + k_{\perp}^2 \lambda_D^2\right) - \nu_i \G_{0i}\right]^2}
\ee
One sees that in the long-wavelength limit, $\G_{0i} \rightarrow 1$, the
universal mode is unstable for finite $k_{\perp} \lambda_D$ 
with $\omega_r$ independent of $\lambda_D$ and $\gamma \sim k_{\perp}^2
\lambda_D^2$ for small $k_{\perp}^2 \lambda_D^2$. 
For large $k_{\perp}^2 \lambda_D^2$, both $\omega_r$ 
and $\gamma \sim 1/k_{\perp}^2 \lambda_D^2$. This behaviour is seen in the
numerical solution of the dispersion relation Eq.~(\ref{disp_rel}) shown in Fig.~\ref{univ_analyt}. 
Here, we use the parameters $k_{\perp} \rho_i = 0.1$, $k_{\|} \rho_i = 7.43 \times 10^{-4}$, 
$\kappa_{ni} \rho_i = \kappa_{ne} \rho_i = \kappa_{np} \rho_i = 0.3$, and $\kappa_{Ti} \rho_i = \kappa_{Te} \rho_i = \kappa_{Tp} \rho_i = 0.0$ with 
$\kappa_{na} = \df \ln n_a / \df \ln x$ and $\kappa_{Ta} = \df \ln T_a / \df \ln x$.

For $\lambda_D = 0$, the universal mode needs $k_{\perp} \rho_i \sim 1$ to be unstable. The
numerical solution corresponding to this case is shown in Fig.~\ref{univ}. The
dispersion relation (\ref{disp_rel}) is solved for the 
parameters $k_{\perp} \rho_i = 2.0$, $k_{\|} \rho_i = 7.4 \times 10^{-4}$,
$\kappa_{{T}i} = \kappa_{{T}e} = 0$, $\lambda_D = 0$. One sees that the
universal instability can exist in pair plasmas in slab geometry but 
requires both the proton fraction and the ion density gradient to exceed 
than some threshold. Practically, it suggests that the universal mode will be
stable in pair plasmas if the proton contamination is small. Interestingly,
the {\it positron} density gradient has zero effect on 
the universal mode if quasineutrality 
$n_e = n_p + n_i$ is assumed since any effect of the positron density gradient
on the universal mode is perfectly cancelled by the electrons. 
%
%=================================================================
%
\section{ITG instability} \label{ITG_mode}
\begin{figure}
  \vspace{1.0cm}
    \centerline{
      \includegraphics[width=7cm]{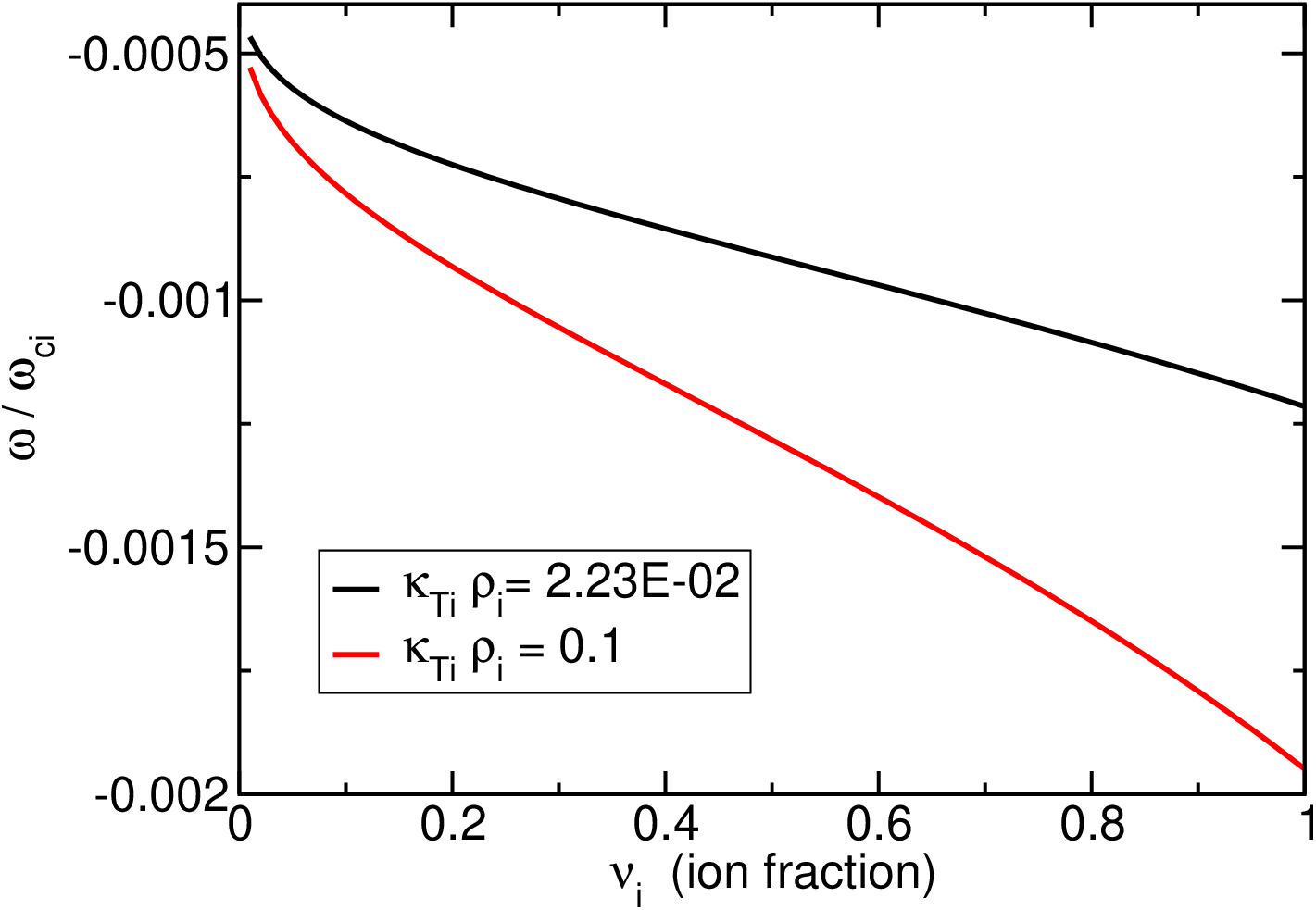}
      \hspace{0.2cm}
      \includegraphics[width=7cm]{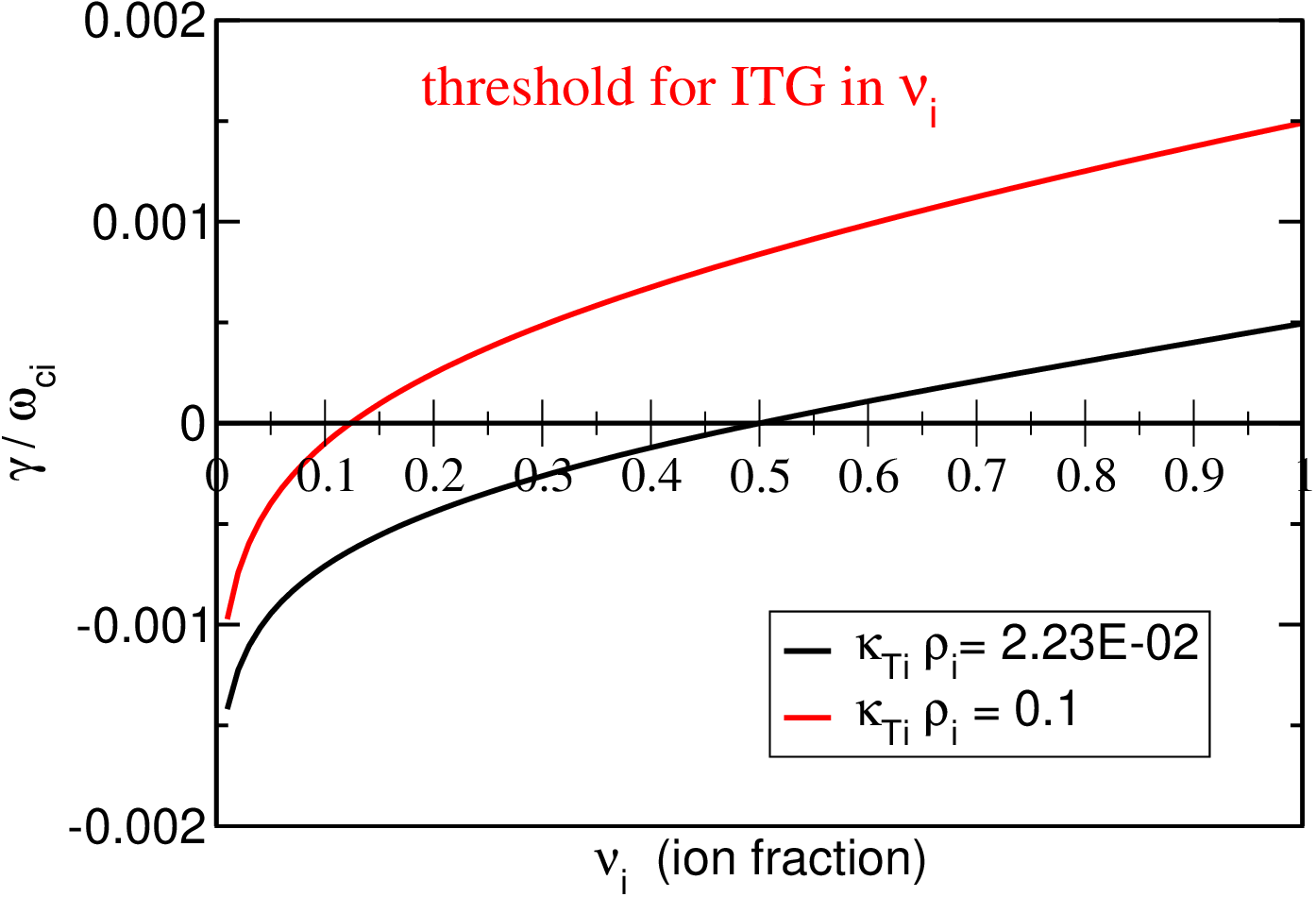} 
    }
\caption{Effect of proton contamination on the ITG mode in a pair plasma. The
  wave numbers are 
$k_{\perp} \rho_i = 0.24$ and $k_{\|} \rho_i = 7.4 \times 10^{-4}$. The density
and the electron temperature profiles are flat, 
$\kappa_{{\rm T}i} = \df \ln T_i(x) / \df \ln x$, and $\tau_i = 1$.}
\label{ITG_nu}
\end{figure}

For simplicity, we consider the flat-density limit. In this case, it is convenient to define 
$\omega_{Ta}=\eta_{a}\omega_{*a}$ with $a = i, e, p$ being the species index. For electrons 
and positrons, we assume flat profiles $\omega_{Te} = \omega_{Tp} = 0$. For
ions, the temperature gradient is finite $\omega_{Ti} \ne 0$. To allow for
unequal temperatures of different species, we introduce the notation:
\be
\label{hat_nu}
\hat{\nu}_a = \frac{2 \, \nu_a/\tau_a}{\sum_{a'} \nu_{a'}/\tau_{a'}}
\ee
with $\nu_a = n_a/n_e$ and $\tau_a = T_a/T_e$. 
Note that quasineutral plasmas satisfy both $\sum_a \nu_a = 2$ and 
$\sum_a \hat{\nu}_a = 2$. If the temperatures of all species 
are equal ($\tau_a = 1$) in such plasmas, then $\hat{\nu_a} = \nu_a$.
In our notation, the dispersion relation becomes 
\be
1 + k_{\perp}^2 \lambda_D^2 \,+\, \sum_{a=i,p,e} \frac{\hat{\nu}_a}{2} \, \zeta_a \, Z_{0a} \G_{0a} + 
\frac{\hat{\nu}_i \omega_{Ti} \, \zeta_i}{2 \, \omega} \left( \frac{3}{2} Z_{0i} \G_{0i}
    - Z_{0i} \G_{*i} - Z_{2i} \G_{0i} \right)  = 0
\ee
We consider the long wave-length limit $\G_{0a} = \G_{*a} = 1$ for all particle
species. For the ITG instability, we can assume 
$k_{\|} v_{{\rm th}i} \ll \omega \ll k_{\|} v_{{\rm th}e}$. Then, the plasma dispersion function can be expanded as
\be
Z_0(\zeta_i) \approx \,-\, \frac{1}{\zeta_i} 
- \frac{1}{2 \zeta_i^3} - \frac{3}{4 \zeta_i^5}
%+ i \sqrt{\pi}  e^{-\zeta_i^2} 
\ , \;\;
Z_0(\zeta_p) = Z_0(\zeta_e) \approx i \sqrt{\pi} %- 2 \zeta_e 
\ee
To leading order, we obtain the dispersion relation
\be
1 - \frac{\hat{\nu}_i}{2} + k_{\perp}^2 \lambda_D^2 = 
\,-\, \frac{\hat{\nu}_i \omega_{Ti}}{4 \omega^3} \, k_{\|}^2 v_{{\rm th}i}^2
\ee
Noting that $\omega_{Ti} < 0$, we find the unstable solution of this
dispersion relation: 
\be
\label{eq:fluid_ITG}
\omega = \frac{1}{2^{1/3}} \left( \frac{\hat{\nu}_{i}|\omega_{Ti}| k_{\|}^{2}v_{{\rm th}i}^{2}}
{ 2 - \hat{\nu}_{i} + 2 k_{\perp}^{2}\lambda_{D}^{2} }\right)^{1/3} 
\left(\,-\,\frac{1}{2} + i\,\frac{\sqrt{3}}{2}\right)
\ee
This root corresponds to the well-known fluid limit of the slab ITG
instability \citep{Coppi_Rosenbluth_Sagdeev}. Note that the ITG frequency is
negative, as expected. One sees that in an ion-contaminated
electron-positron plasma, the frequency and growth rate of the fluid ITG
instability are proportional to 
$\Big(\hat{\nu}_i |\omega_{Ti}|\Big)^{1/3}$. Hence, pure pair plasmas with
$\hat{\nu}_i = 0$ cannot support the slab ITG. Similarly to the frequency and
the growth rate, the destabilisation threshold is also determined by the
product $\hat{\nu}_i |\omega_{Ti}|$, and not just $|\omega_{Ti}|$ as is the
case for conventional (e.~g.~hydrogen) plasmas. Numerical results 
demonstrating this prediction are shown in Fig.~\ref{ITG_nu}.  
Here, the dependence of the ITG frequency and the growth rate
on the proton contamination is plotted. One sees that the absolute value of
the frequency indeed decreases strongly at a smaller proton content, in
agreement with the analytic result. One also sees that the mode is
unstable only when the proton  content exceeds some threshold, whose value
depends on the ion temperature gradient. This is of practical interest since
it indicates that the ITG modes may be stable at a large ion temperature
gradient in ion-contaminated pair plasmas if the ion fraction is small enough.

\begin{figure}
  \vspace{1.0cm}
    \centerline{
      %\includegraphics[width=7cm]{../graph/DISPERS/slab/itg_kTi.eps}
      %\hspace{0.2cm}
      %\includegraphics[width=7cm]{../graph/DISPERS/slab/itg_ky.eps} 
      \includegraphics[width=7cm]{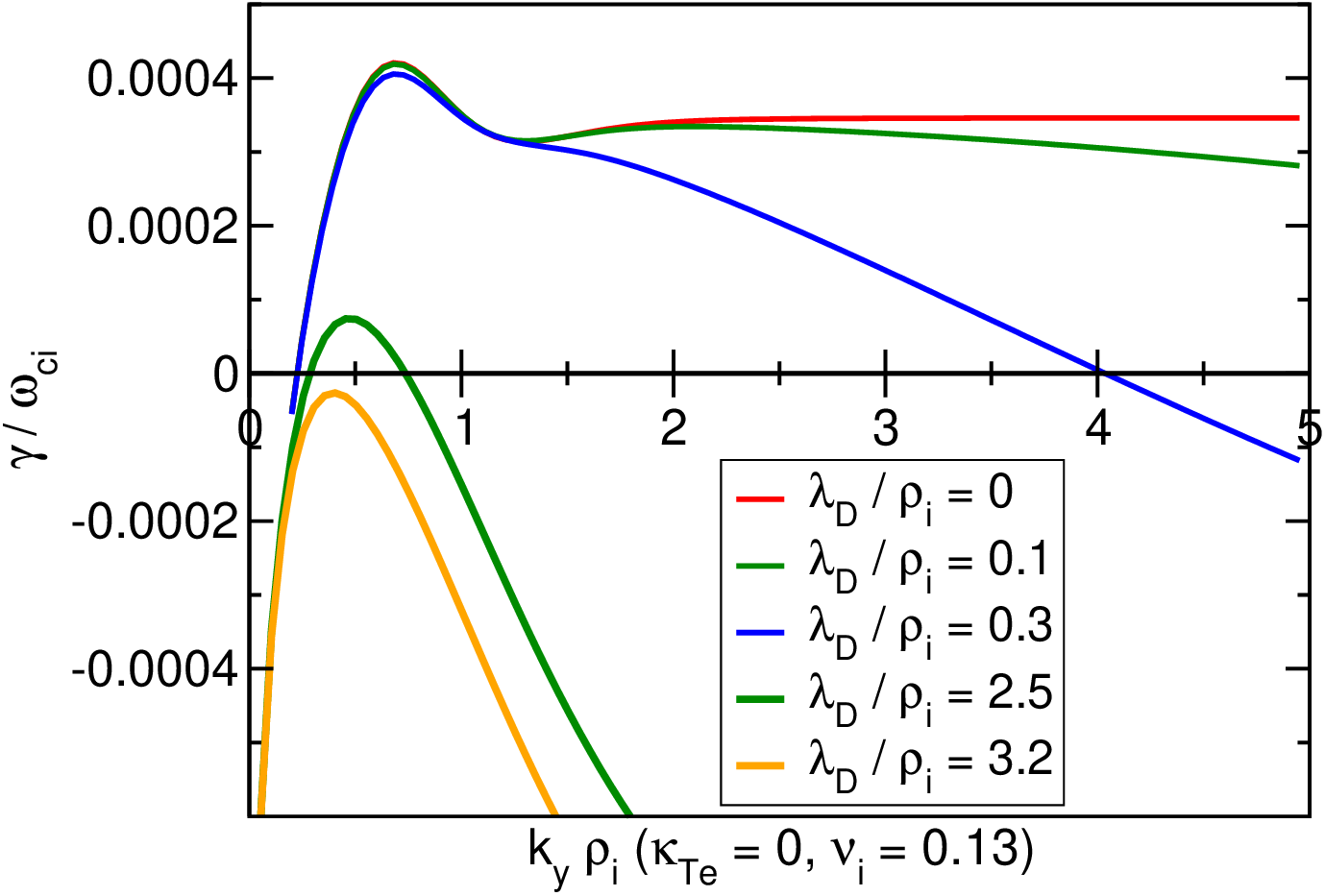} 
    }
\caption{ITG mode in a pair plasma with the proton contamination 
  $\nu_i = 0.13$ for $\tau_i = 1$ and $k_{\|} \rho_i = 7.4 \times
  10^{-4}$. Effects of the finite Debye length is considered.} 
\label{ITG_LD}
\end{figure}

Another aspect of practical interest for the pair-plasma experiment 
\citep{pedersen2012} is the effect of the Debye length on the
microinstabilities. This effect is usually negligible for tokamak or
stellarator plasmas, where the Debye length is much smaller than the ion gyro-radius. 
In the pair-plasma experiment, however, the Debye length is not expected to be
very small and can become comparable to the proton gyroradius. This can have a
strongly stabilising effect on the ITG stability, as shown in
Fig.~\ref{ITG_LD}. One sees that for a given $k_{\|}$, the ITG instability can
disappear for all perpendicular wavelengths if $\lambda_D/\rho_i$ is large enough. 
%
%=================================================================
%
\section{ETG instability} \label{ETG_mode}
\begin{figure}
   \vspace{1.0cm}
     \centerline{
       \includegraphics[width=7cm]{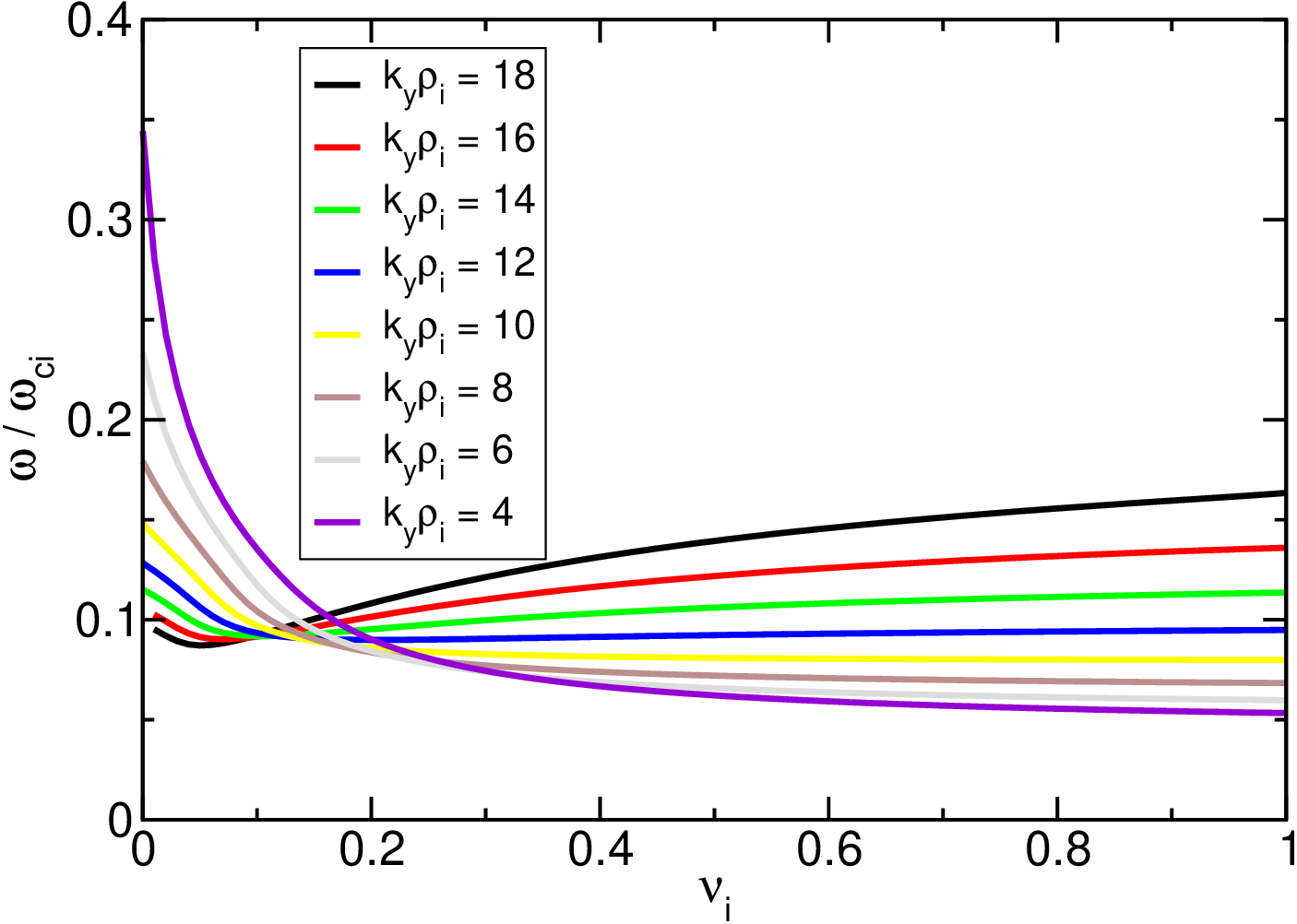}
       \hspace{0.5cm}
       \includegraphics[width=7cm]{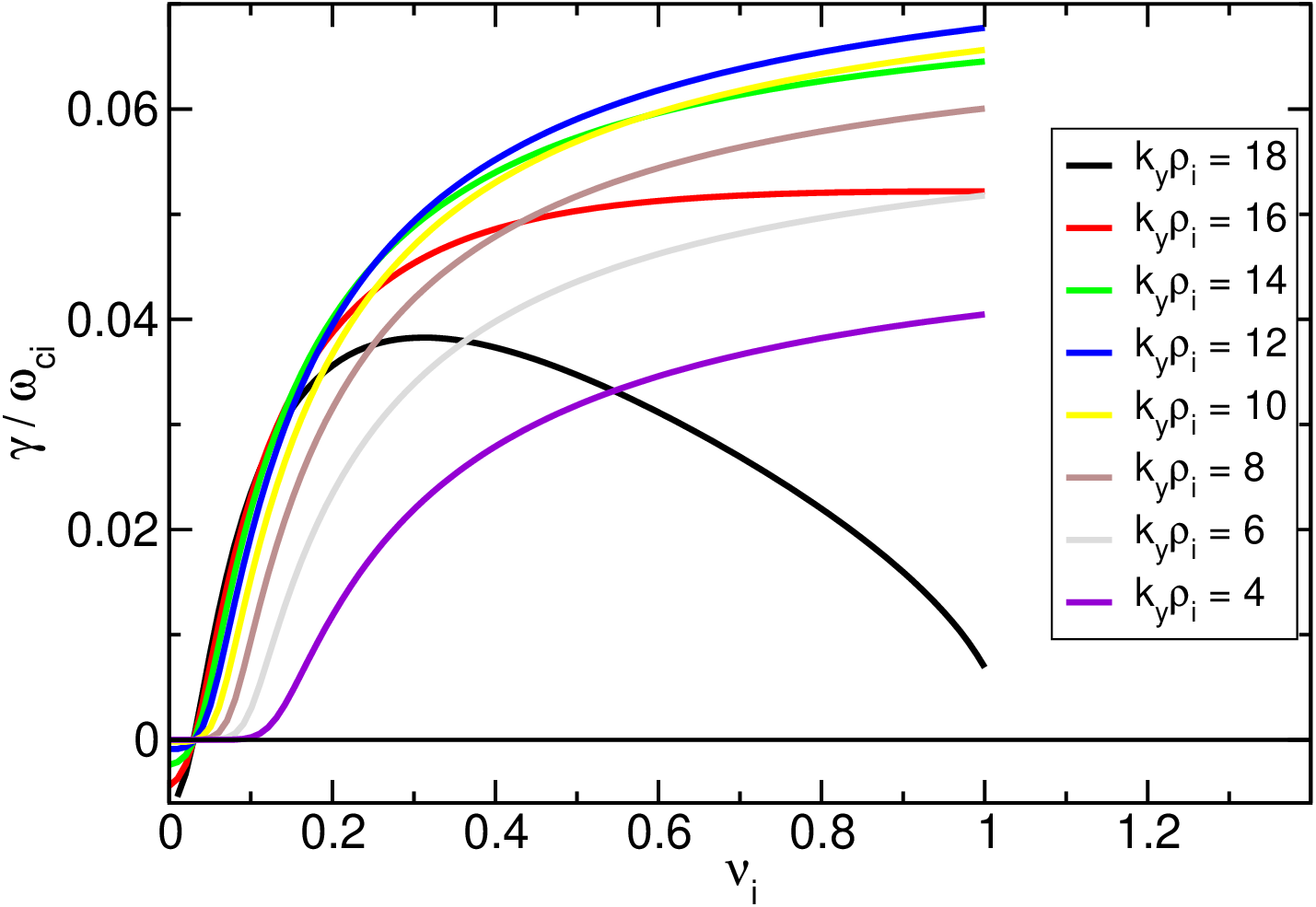} 
     }
 \caption{Frequency and growth rate of the ETG mode in a 
   three-component electron-positron-proton plasma for $\omega_{Tp} =
   \omega_{Te}$. One sees that the ion fraction must exceed some
   threshold for the ETG to be unstable. Here, 
   $k_{\|} \rho_i = 7.4 \times 10^{-4}$, $\kappa_{{\rm n}i} = 0$, $\lambda_D =
   0$, and $\tau_a = 1$.}
 \label{ETG}
 \end{figure}

Consider now the case when only electron and positron temperature
gradients are present, i.~e. $\omega_{T(e,p)} \ne 0$, while $\omega_{Ti} = 0$
and $\omega_{*(e,p,i)} = 0$ (flat density). In this Section, we will also allow for 
unequal temperatures of different species. Therefore, the notation defined in
Eq.~(\ref{hat_nu}) will be used. In this notation, the dispersion relation is  
\be
1 + k_{\perp}^2 \lambda_D^2 + \sum_{a=p,e} \frac{\hat{\nu}_a}{2} \, \zeta_a \,
\left[Z_{0a} \G_{0a} + \frac{\omega_{Ta}}{\omega} \left( \frac{3}{2} Z_{0a} \G_{0a}
    - Z_{0a} \G_{*a} - Z_{2a} \G_{0a} \right) \right] = 0
\ee
For the perpendicular wave numbers, we assume $k_{\perp} \rho_i \gg 1$ but $k_{\perp} \rho_{(e,p)} \ll 1$ . Then
\be
\G_{0i} = 0 \ , \;\; \G_{*i} = 0 \ , \;\;
\G_{0(e,p)} = 1  \ , \;\; \G_{*(e,p)} = 1 
\ee
Assuming large frequencies $\omega \gg k_{\|} v_{{\rm th}(e,p)}$, we can write
\be
Z_0(\zeta_{e,p}) \approx \,-\, \frac{1}{\zeta_{e,p}} - \frac{1}{2 \zeta_{e,p}^3} - \frac{3}{4 \zeta_{e,p}^5}  
\ee
Under these assumptions, the dispersion relation reduces in the leading order to 
\ba
\left( \frac{\hat{\nu}_i}{2} + k_{\perp}^2 \lambda_D^2 \right) + 
\frac{\hat{\nu}_e \tau_e \omega_{Te} + \hat{\nu}_p \tau_p \omega_{Tp}}{4 \omega \zeta_e^2} = 0
\ea
Here, the relations $\sum_a \hat{\nu}_a = 2$ %with $a=(i,e,p)$ 
and $\zeta_p^2 = \zeta_e^2 / \tau_p$ have been employed. Let us now consider
the case of equal electron and positron 
temperature profiles, implying $\tau_p = \tau_e = 1$ and 
$\omega_{Te} + \omega_{Tp} = 0$. Recall that $\tau_a = T_a/T_e$, and our sign
conventions imply $\omega_{Te} > 0$ and $\omega_{Tp} < 0$. These conditions
are likely since the characteristic time of the energy exchange between the
electrons and the positrons is comparable to their Maxwellisation time. If the
plasma has had time to reach a locally Maxwellian state, as we have assumed,
the electron and positron temperatures should also have equilibrated. 
Then, the unstable solution is  
\be
\label{eq:fluid_ETG}
\omega = \frac{1}{2^{1/3}} \left( \frac{k_{\|}^{2}v_{{\rm th}e}^{2}}{
      \hat{\nu_{i}} + 2 k_{\perp}^{2}\lambda_{D}^{2} }
\; \hat{\nu}_{i} \tau_i \omega_{Te} \right)^{1/3} 
\left( \frac{1}{2} + i\,\frac{\sqrt{3}}{2}\right)
\ee
This solution corresponds to the fluid limit of the slab ETG instability, 
which is similar to the ITG instability, Eq.~(\ref{eq:fluid_ITG}), but has a
frequency of the opposite sign. 
The mode is expected to be stable in a pure pair plasma $\hat{\nu_i} = 0$,
as can indeed be seen from the numerical solution of the full dispersion
relation Eq.~(\ref{disp_rel}), shown in Fig.~\ref{ETG}. In
Eq.~(\ref{eq:fluid_ETG}), however, also the denominator vanishes at
$\hat{\nu_i} = 0$ if $k_{\perp} \lambda_D =0$. This indicates that
higher-order terms must be considered in order to address this limit. Also, it
indicates sensitivity of the ETG mode in contaminated pair plasmas to 
finite-Debye-length effects. 

%% \begin{figure}
%%   \vspace{1.0cm}
%%     \centerline{
%%       \includegraphics[width=7cm]{../graph/DISPERS/slab/etg_nu_w.eps}
%%       \hspace{0.5cm}
%%       \includegraphics[width=7cm]{../graph/DISPERS/slab/etg_nu.eps} 
%%     }
%% \caption{The frequency and the growth rate of the ETG mode in the
%%   three-component electron-positron-proton plasma for $\omega_{Tp} = 0$. One sees that the
%%   positrons may have a strong stabilising effect on the ETG mode, depending on
%%   the perpendicular wave number.}
%% \label{ETG}
%% \end{figure}

\begin{figure}
  \vspace{1.0cm}
    \centerline{
      \includegraphics[width=7cm]{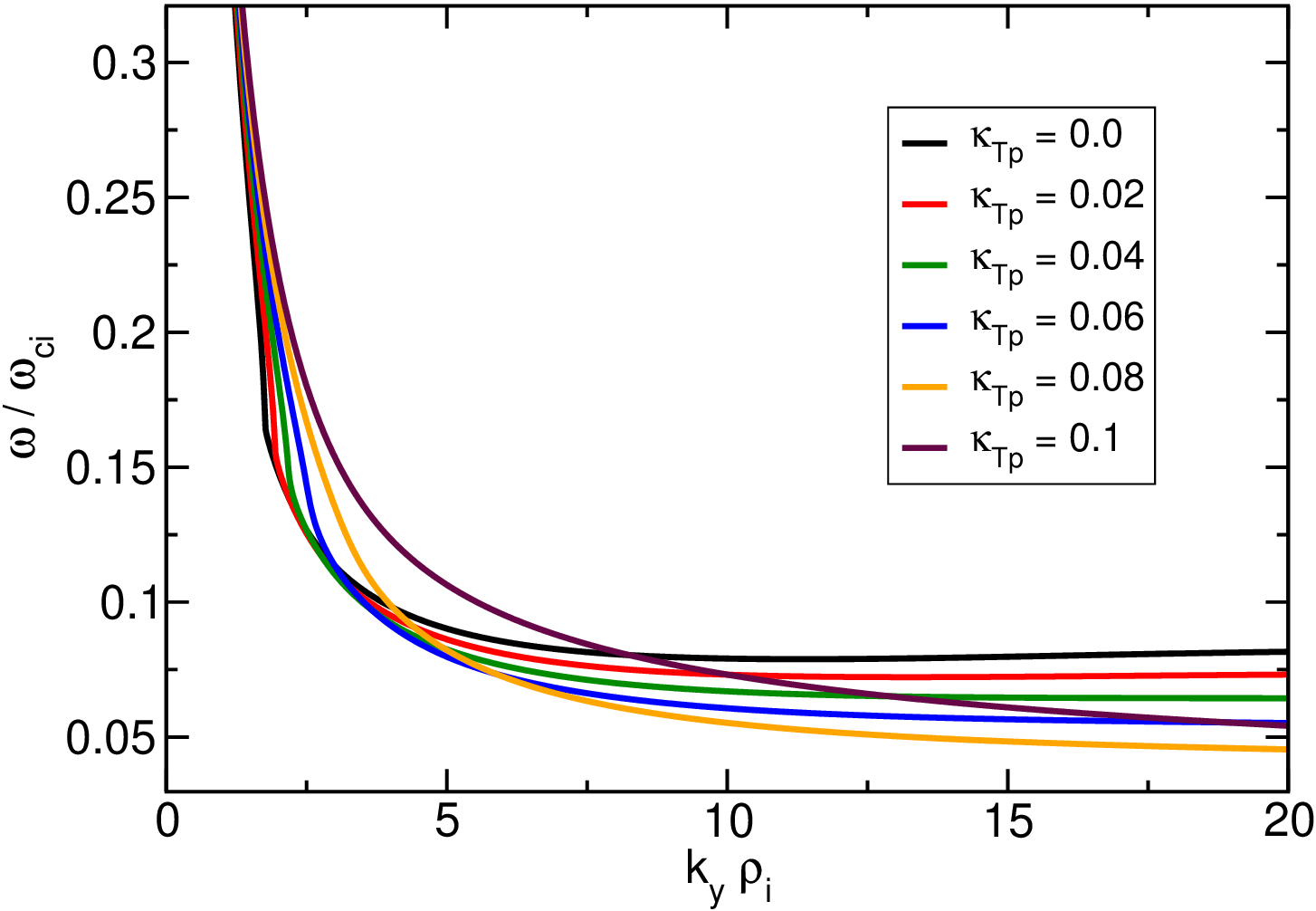}
      \hspace{0.5cm}
      \includegraphics[width=7cm]{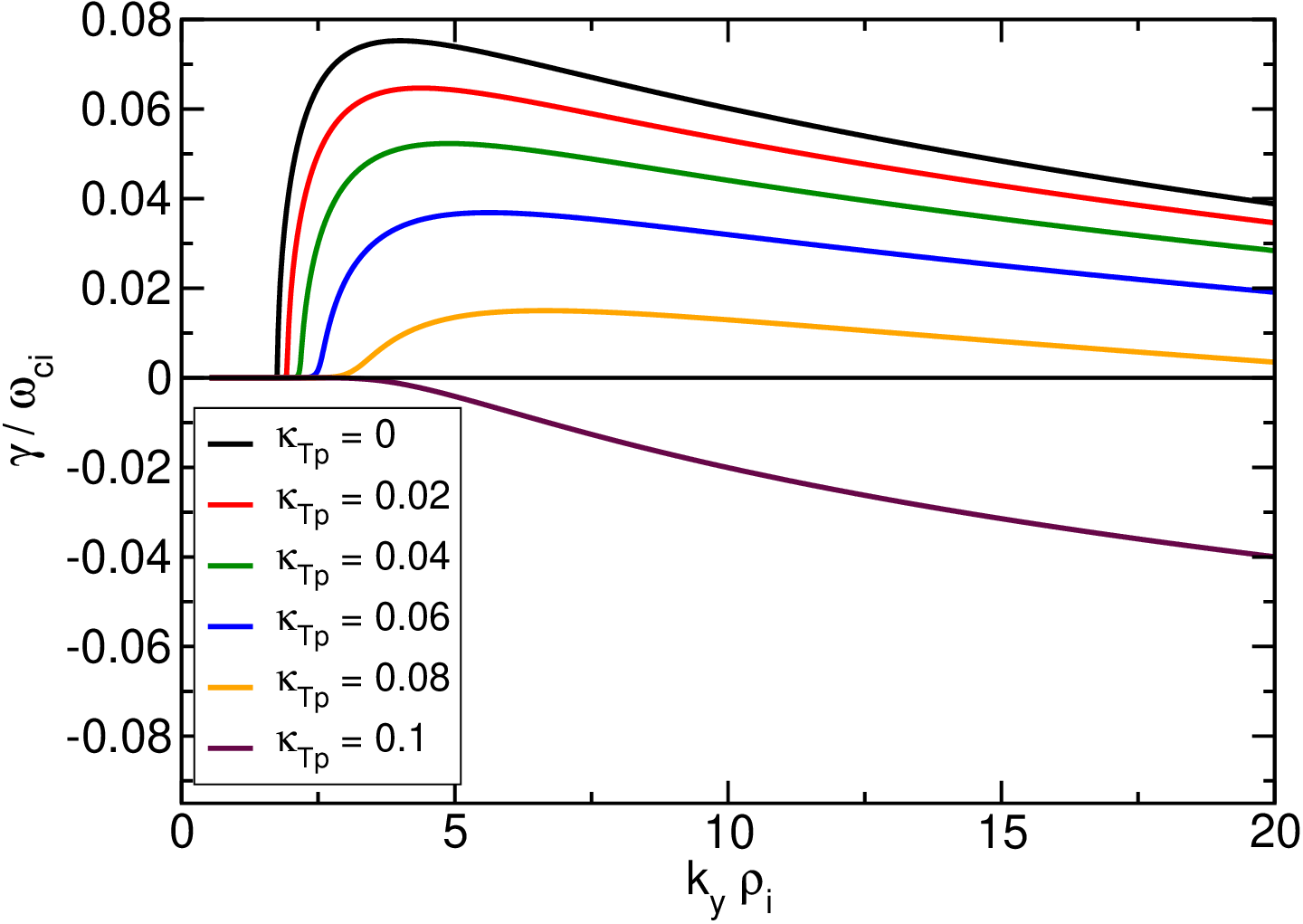} 
    }
\caption{Frequency and growth rate of the ETG mode in a pure pair plasma
  when the symmetry between the species is broken by a difference in the
  electron and positron temperature profiles. The electron temperature
  profile with $\kp_{{\rm T}e} = 0.1$ is kept fixed. Here, $\lambda_D/\rho_i = 0.1$ and $\tau_a = 1$.} 
\label{ETG_pair}
\end{figure}

\begin{figure}
  \vspace{1.0cm}
    \centerline{
      \includegraphics[width=7cm]{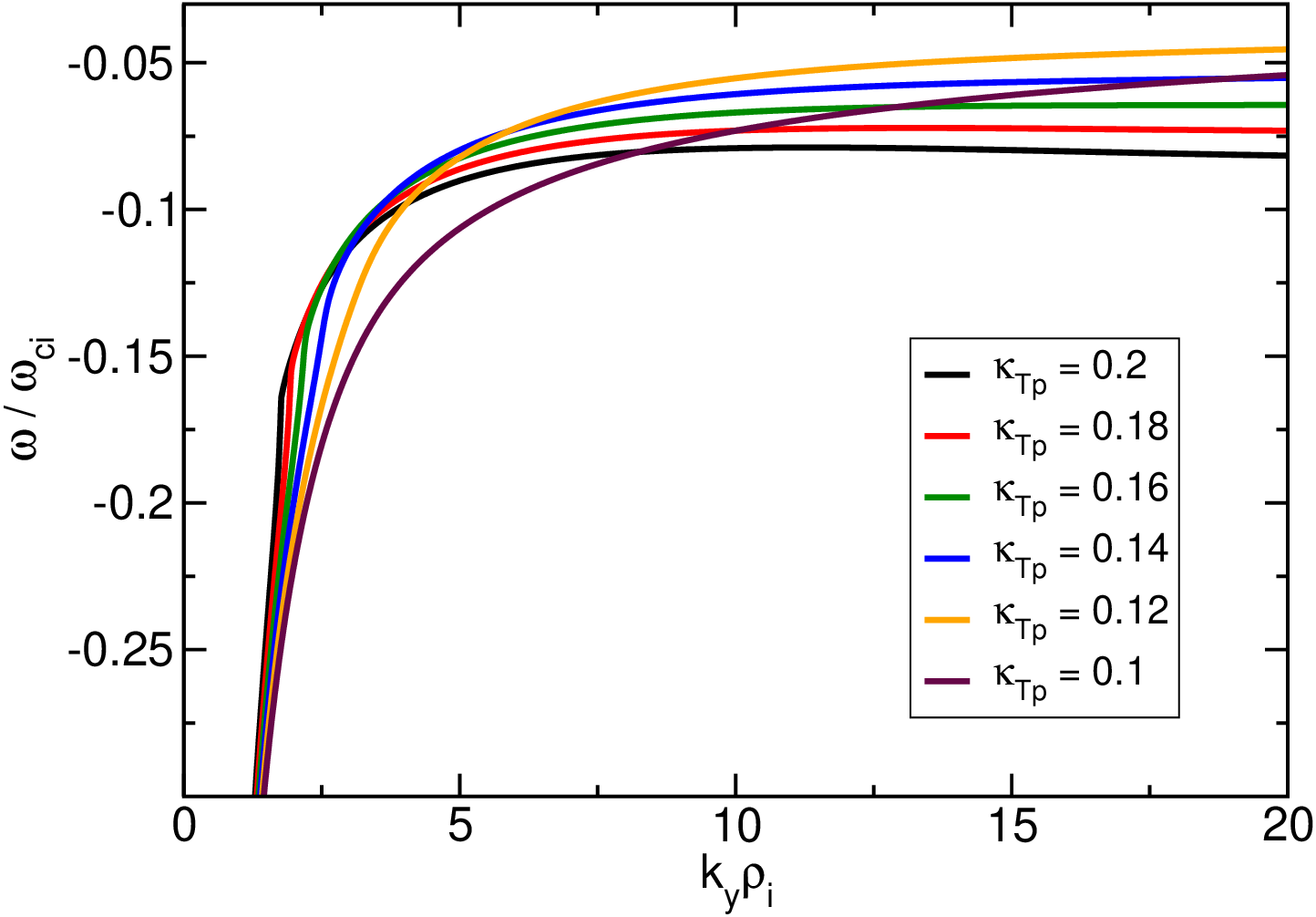}
      \hspace{0.5cm}
      \includegraphics[width=7cm]{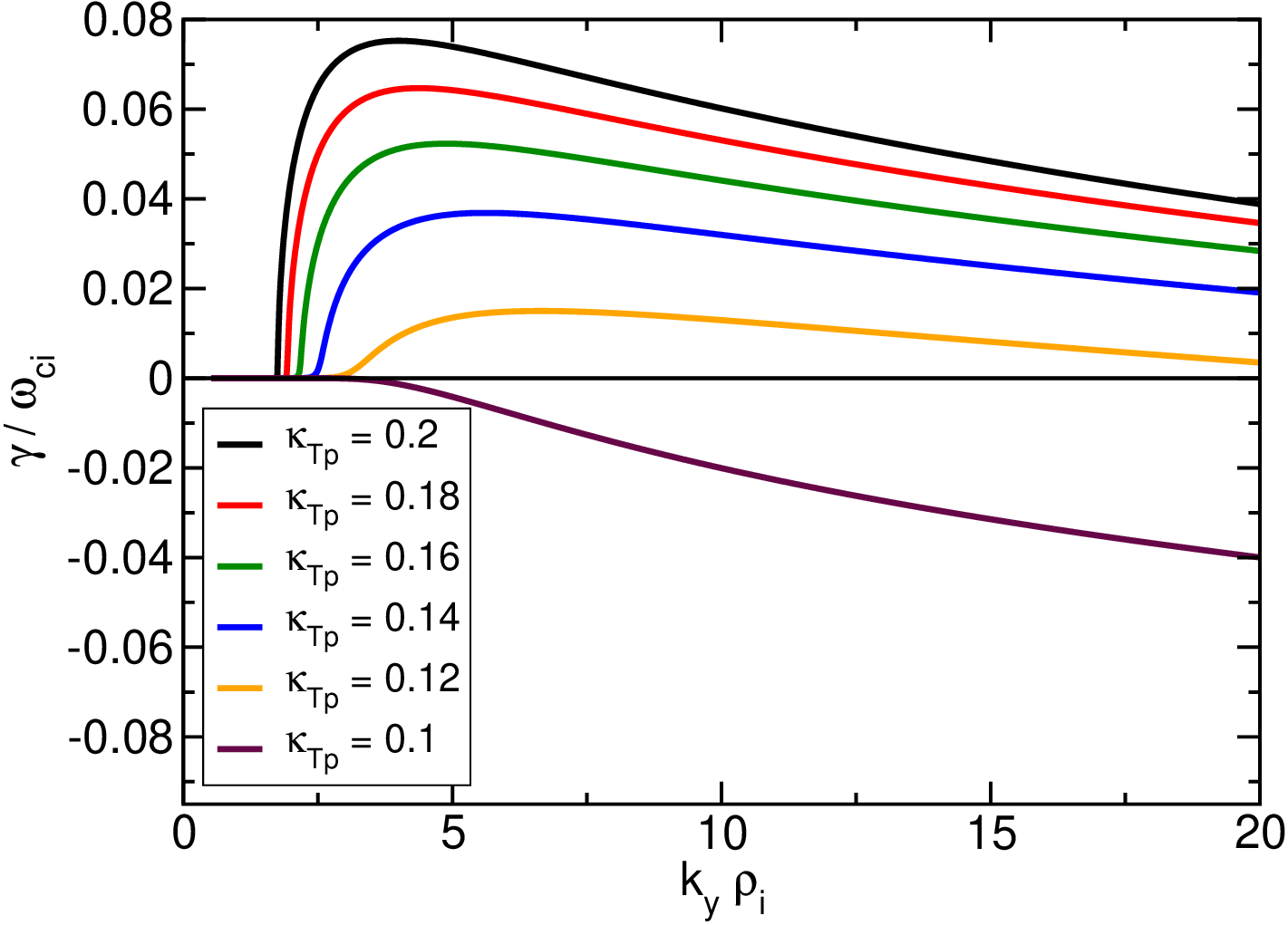} 
    }
\caption{Frequency and growth rate of the PTG mode in a pure pair plasma
  when the symmetry between the species is broken by a difference in the
  electron and positron temperature profiles. The electron temperature
  profile with $\kp_{{\rm T}e} = 0.1$ is kept fixed. Here, $\lambda_D/\rho_i = 0.1$ and $\tau_a = 1$.} 
\label{PTG_pair}
\end{figure}

Interestingly, the ETG mode can be unstable also in a pure pair plasma
($\hat{\nu}_i = 0$) when the
electron and the positron temperature gradients are different for some
reason. Assuming for simplicity $k_{\perp} \lambda_D$ to be finite, we can 
write the unstable ETG solution as 
\be
\label{eq:fluid_ETG1}
\omega = \frac{1}{2^{1/3}} \left( \frac{k_{\|}^{2}v_{{\rm th}e}^{2}}{
    k_{\perp}^{2}\lambda_{D}^{2}} \, \frac{\tau_e \tau_p}{\tau_e + \tau_p} \;  
  \Big[ |\omega_{Te}| - |\omega_{Tp}|\Big] \right)^{1/3}  
\left( \frac{1}{2} + i\,\frac{\sqrt{3}}{2}\right)
\ee
The numerical solution of the dispersion relation Eq.~(\ref{disp_rel})
corresponding to a pure pair plasma ETG is shown in Fig.~\ref{ETG_pair}.  
This result is valid if the electrons have a steeper temperature profile.
% or are colder than the positrons (implying $\tau_p > \tau_e$). 
Otherwise, the ETG
instability is replaced by the PTG (Positron Temperature Gradient driven)
instability, which has a negative frequency: 
\be
\label{eq:fluid_PTG}
\omega = \frac{1}{2^{1/3}} \left( \frac{k_{\|}^{2}v_{{\rm th}e}^{2}}{
    k_{\perp}^{2}\lambda_{D}^{2}} \, \frac{\tau_e \tau_p}{\tau_e + \tau_p} \; 
  \Big[ |\omega_{Tp}| - |\omega_{Te}|\Big]\right)^{1/3}  
\left( \,-\,\frac{1}{2} + i\,\frac{\sqrt{3}}{2}\right)
\ee
The PTG solution is shown in Fig.~\ref{PTG_pair}. 
%
%=================================================================
%
\section{Shear Alfv\'en wave} \label{alfven_wave}
\begin{figure}
  \vspace{1.0cm}
    \centerline{
      \includegraphics[width=7cm]{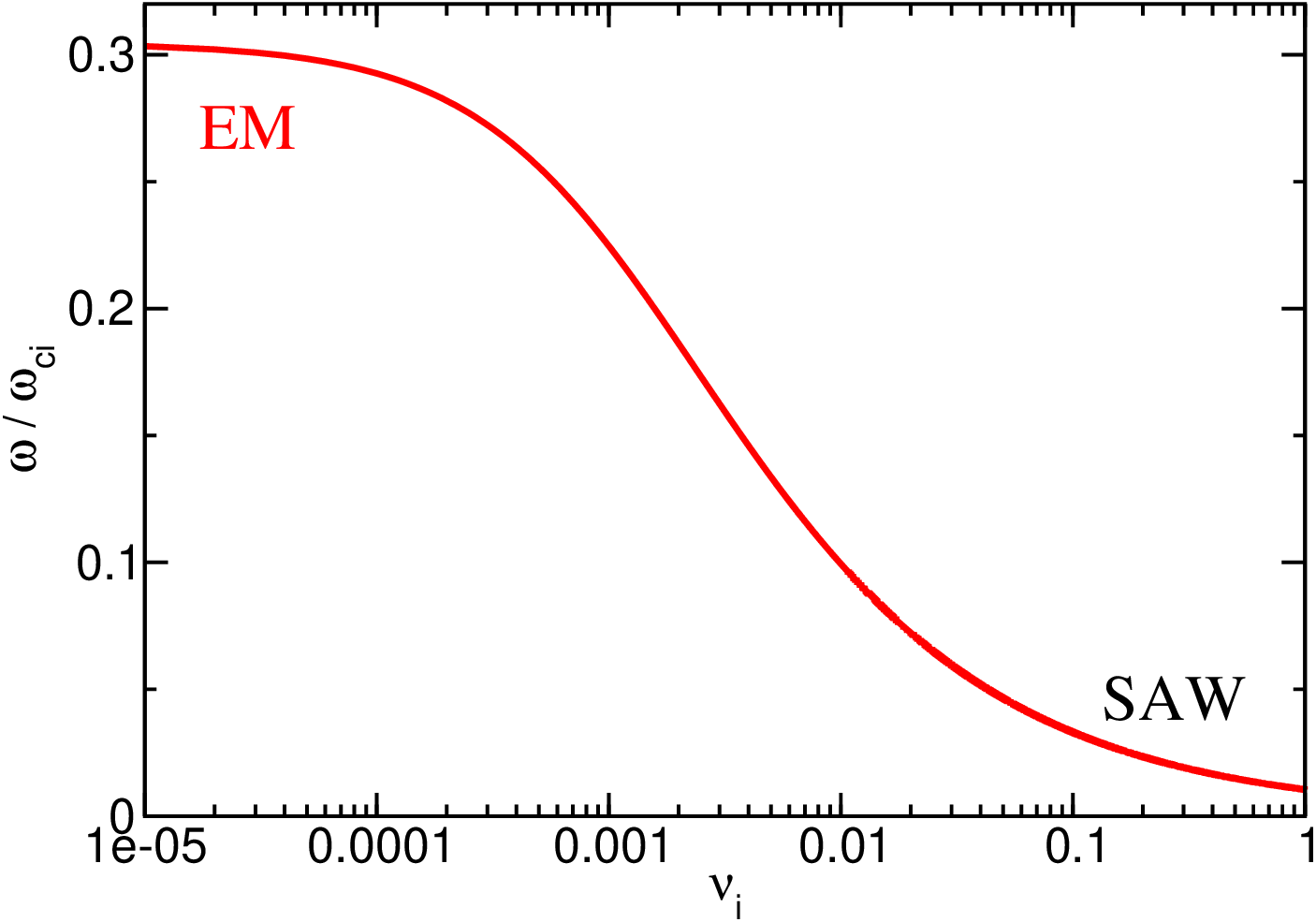}
      \hspace{0.2cm}
      \includegraphics[width=7cm]{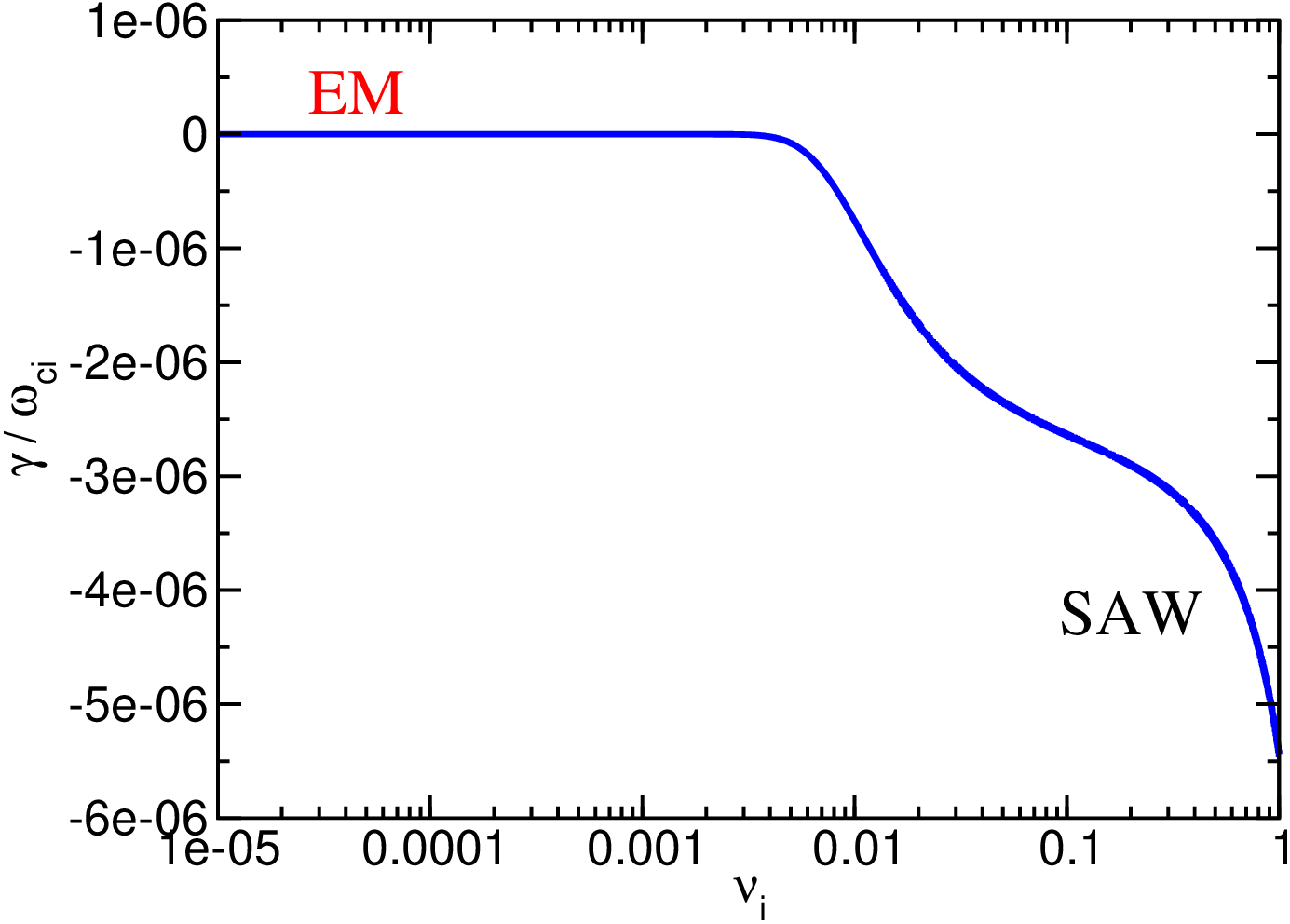} 
    }
% \caption{Frequency and damping rate of the shear Alfv\'en wave as a function
%   of the ion contamination in the pair plasma for %. The parameters %used are 
%   $k_{\perp} \rho_i = 0.475$, $k_{\|} \rho_i = 7.4 \times 10^{-4}$, $\lambda_D
%   = 0$, and 
%   $\beta_e = 0.005$.}
    \caption{Frequency and growth rate of the shear Alfv\'en wave (SAW) as a
      function of ion contamination in a pair plasma for $k_{\perp}
      \rho_i = 0.05$, $k_{\|} \rho_i = 7.4 \times 10^{-4}$, $\lambda_D/\rho_i = 0.01$,
      and $\beta_e = 0.005$. One sees the transition from the SAW regime
      $\nu_i \sim 1$ to a regime of an electromagnetic wave travelling at the speed of light
      when $\nu_i \rightarrow 0$. The latter limit is not properly described
      by the gyrokinetic theory of this paper since the relativistic effects must
      be taken into account in the wave dynamics \citep{Zocco_alf}.} 
\label{alf}
\end{figure}

Finally, we consider a homogeneous plasma (all profiles are flat) and solve
the electromagnetic dispersion relation Eq.~(\ref{disp_rel}) at finite
$\beta$.  
Assuming $k_{\|} v_{{\rm th}i} \ll \omega \ll k_{\|} v_{{\rm th}e}$, one can write: 
\ba
&&{} Z_{0i} = \,-\,\frac{1}{\zeta_i} - \frac{1}{2 \zeta_i^3} %+ i \sqrt{\pi} e^{-\zeta_i^2} 
+ {\cal O}\left(\frac{1}{\zeta_i^5}\right) 
\ , \;\; Z_{0(e,p)} = i \sqrt{\pi} + {\cal O}(\zeta_e) \\
&&{} Z_{1i} = - \frac{1}{2 \zeta_i^2}
+ {\cal O}\left(\frac{1}{\zeta_i^4}\right) 
\ , \;\; Z_{1(e,p)} = 1 + {\cal O}(\zeta_e) \\
&&{} Z_{2i} = \,-\,\frac{1}{2 \zeta_i} + {\cal O}\left(\frac{1}{\zeta_i^3}\right) 
\ , \;\; Z_{2(e,p)} = \zeta_e + {\cal O}(\zeta_e^2)
\ea
For flat profiles $\omega_{*a} = 0$ and $\eta_a = 0$. Hence, from Eq.~(\ref{I_na_def2}) %and (\ref{I_na_def2})
\be
W_{0a} = \zeta_a Z_{0a} \G_{0a} \ , \;\;
W_{1a} = \zeta_a Z_{1a} \G_{0a} \ , \;\;
W_{2a} = \zeta_a Z_{2a} \G_{0a} 
\ee
Employing the appropriate expansions of the plasma dispersion function, we obtain:
\ba
&&{} W_{0i} = \,-\, \G_{0i} - \frac{\G_{0i}}{2 \zeta_i^2} + {\cal O}\left(\frac{1}{\zeta_i^4}\right)  \ , \;\;
W_{0(e,p)} = i \zeta_{e,p} \sqrt{\pi} + {\cal O}\left(\zeta_{e,p}^2\right) \\
&&{} W_{1i} =   \,-\, \frac{\G_{0i}}{2 \zeta_i}
+ {\cal O}\left(\frac{1}{\zeta_i^2}\right) \ , \;\;
W_{1(e,p)} = \zeta_{e,p} + {\cal O}\left(\zeta_{e,p}^2\right) \\
&& W_{2i} = \,-\,\frac{\G_{0i}}{2} + {\cal O}\left(\frac{1}{\zeta_i^2}\right) \ , \;\;
W_{2(e,p)} = \zeta_{e,p}^2 + {\cal O}\left(\zeta_{e,p}^3\right)
\ea
For equal temperatures and charges of the species, Eq.~(\ref{disp_rel}) becomes
\ba
&&{} \left(1 + k_{\perp}^2 \lambda_D^2 + \frac{1}{2}\sum_a \nu_a W_{0a} \right)
\left(1 - \sum_a \frac{2 \beta_a}{k_{\perp}^2 \rho_a^2} W_{2a}\right) \,+\, \\
&&{} \,+\, \sum_a \nu_a \; W_{1a} v_{{\rm th}a} \;
\sum_a \frac{\beta_a}{k_{\perp}^2 \rho_a^2} \frac{W_{1a}}{v_{{\rm th}a}} = 0 \nonumber
\ea
Here, the notation $\beta_a = \mu_0 n_a T_a / B^2$ is used and the usual
assumption $\beta_a \ll 1$ is made. We will substitute the approximate expressions for $W_{na}$, derived above, 
into this dispersion relation. Note that small terms of the order
$1/\zeta_i^2$ and $1/\zeta_i$ must be kept in $W_{0i}$ and $W_{1i}$,
respectively, since they give order unity contributions in the dispersion
relation when multiplied with $\zeta_e^2$ appearing in $W_{2(e,p)}$ and
$W_{1(e,p)}^2$.  
For equal temperatures and charges of the species, one can write:
\be
\frac{\zeta_e^2}{\rho_e^2} = \frac{\zeta_i^2}{\rho_i^2} \ , \;\;
\zeta_e v_{{\rm th}e} = \zeta_i v_{{\rm th}i}  \ , \;\;
\beta_a = \nu_a \beta_e \ , \;\; 
\nu_i + \nu_p = 1 
\ee
Using these relations and assuming $k_{\|} v_{{\rm th}i} \ll \omega \ll k_{\|} v_{{\rm th}e}$, one can write the dispersion 
relation to the lowest order as follows:
\ba
&&{} \left( 1  + k_{\perp}^2 \lambda_D^2 - \frac{\nu_i \G_{0i}}{2} \right) \left[ 1 - \frac{2 \beta_e
    \zeta_e^2 (1 + \nu_p)}{k_{\perp}^2 \rho_e^2} \right] + 
(1 + \nu_p)^2 \frac{\beta_e \zeta_e^2}{k_{\perp}^2 \rho_e^2}  \,+\, \\
&&{} \,+\, \frac{\nu_i \beta_e \G_{0i}}{k_{\perp}^2 \rho_i^2}  \left[\frac{\nu_i}{2} \left(1 - \G_{0i}\right) + k_{\perp}^2 \lambda_D^2 \right]= 0 \nonumber
\ea
For conventional plasmas with $\nu_i = 1$ and $\nu_p = 0$, and the
long-wavelength approximation for $\G_{0i}$, this dispersion relation reduces
to the shear Alfv\'en wave (SAW):
\be
2 \beta_e \zeta_i^2 = 1 \;\; \Leftrightarrow \;\; 
\omega^2 = k_{\|}^2 \; \frac{B^2}{\mu_0 m_i n_{0e}} = k_{\|}^2 v_A^2 \equiv \omega_A^2
\ee
For a finite positron fraction, one can write
\be
2 \beta_e \zeta_i^2 = \frac{1}{\nu_i} \; \frac{2 - \nu_i \G_{0i} + {\cal O}(\beta_e)}{2 - \nu_i} \; 
\frac{k_{\perp}^2 \rho_i^2}{1 - \G_{0i}}
\ee
if the Debye length is neglected. In the long-wavelength approximation
\be
\label{w_Ai}
2 \beta_e \zeta_i^2 = \frac{1}{\nu_i} \;\; \Leftrightarrow \;\;
\omega = \omega_A / \sqrt{\nu_i} =  k_{\|} \; \frac{B}{\sqrt{\mu_0 m_i n_{0i}}} = k_{\|}^2 v_{Ai}^2 \equiv \omega_{Ai}
\ee
The numerical solution of the full dispersion relation Eq.~(\ref{disp_rel}) for the shear Alfv\'en wave parameters is shown in Fig.~\ref{alf}.
One sees, as expected, that the frequency of the shear Alfv\'en wave
increases very rapidly when $\nu_i \rightarrow 0$ (note the logarithmic scale
in the Figure), in agreement with our findings and \citet{per_jpp}. 

\begin{figure}
  \vspace{1.0cm}
    \centerline{
      \includegraphics[width=7cm]{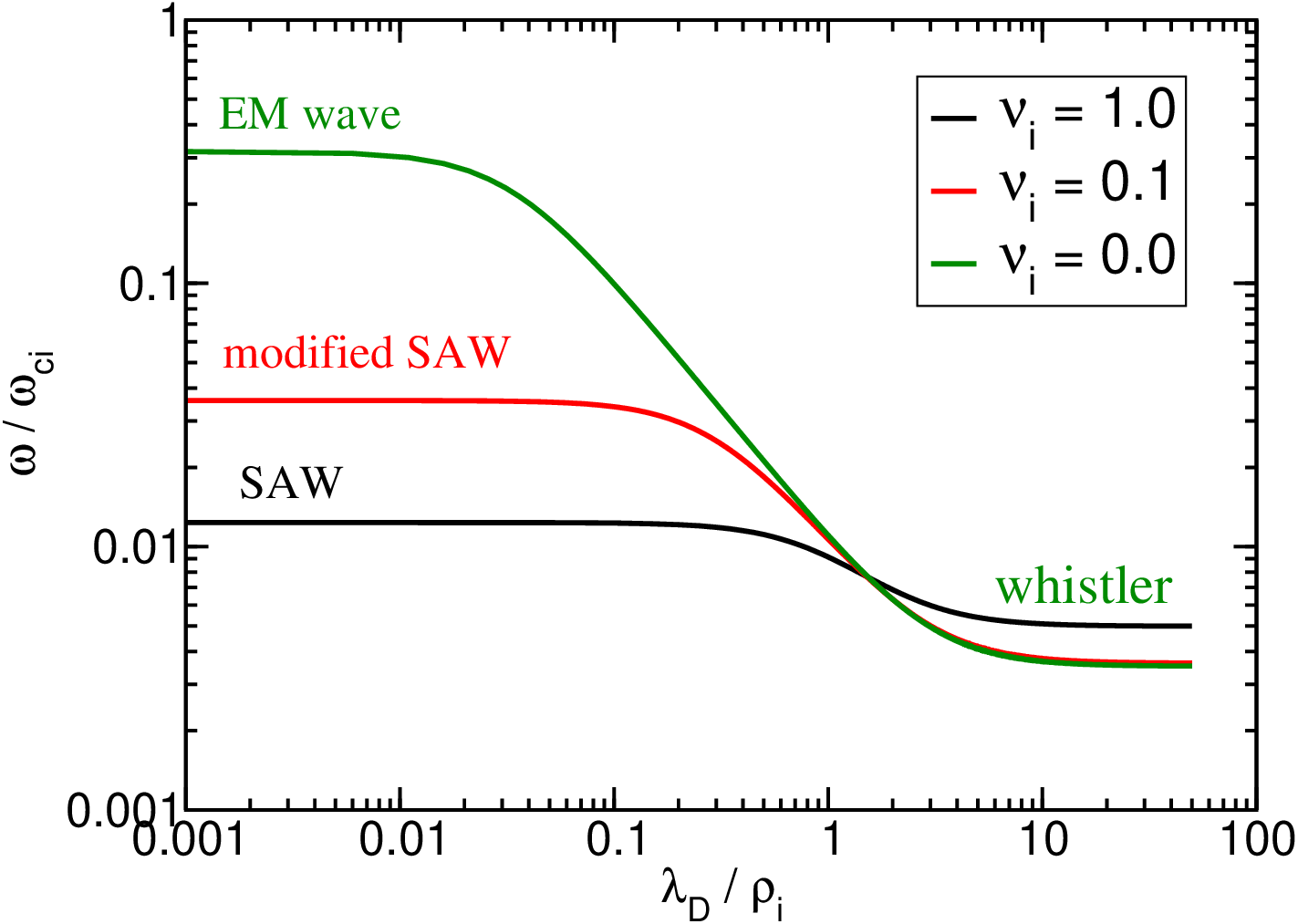}
      \hspace{0.2cm}
      \includegraphics[width=7cm]{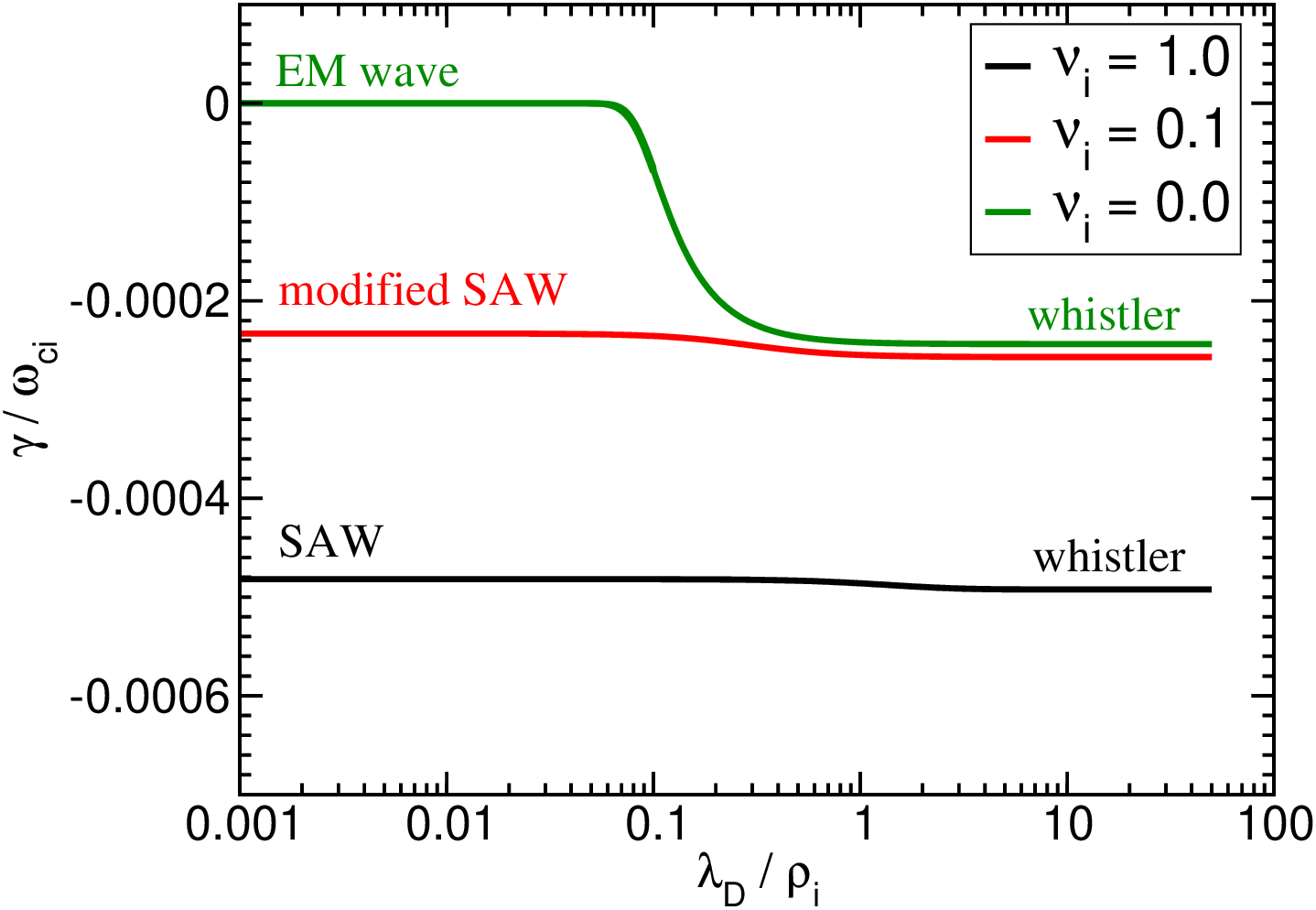} 
    }
\caption{Frequency and growth rate of the shear Alfv\'en wave (SAW),
  ``whistler'' and electromagnetic (EM) wave as a function
  of the Debye length in a conventional plasma, proton-contaminated pair plasma and pure pair plasma. 
  Transitions between different regimes are clearly seen. The parameters used
  are $k_{\perp} \rho_i = 0.475$, $k_{\|} \rho_i = 7.4 \times 10^{-4}$, and 
  $\beta_e = 0.005$. Note that $\lambda_D \gtrsim \rho_e/\sqrt{\beta_e}$
  implies $v_{{\rm th}e} \gtrsim c$. This case is not properly described by
  the gyrokinetic theory of this paper since relativistic effects must be taken into
 account in the particle dynamics \citep{Zocco_alf}.}
\label{alf_LD}
\end{figure}

Note that Eq.~(\ref{w_Ai}) highlights the role of the ions, which carry most of the plasma inertia even at small $\nu_i$, but 
it is singular for $\nu_i = 0$. This formal singularity can be resolved taking
the finite Debye length into account. 
In the long-wavelength approximation %, one obtains:
%and assuming $k_{\perp} \lambda_D \ll 1$, this gives
\be
\label{alf_LD:eq}
2 \beta_e \zeta_i^2 = \frac{2 - \nu_i + 2 k_{\perp}^2 \lambda_D^2}{\nu_i k_{\perp}^2 \rho_i^2  + 2 k_{\perp}^2 \lambda_D^2} \;
\frac{k_{\perp}^2 \rho_i^2}{2 - \nu_i}
= \frac{1}{\nu_i + 2\lambda_D^2 / \rho_i^2} \left[ 1 + {\cal O}\left(k_{\perp}^2 \lambda_D^2\right) \right]
\ee
This equation describes coupling of the ``ion shear Alfv\'en wave'', based on 
ion inertia, to a wave travelling at the speed of light \citep{Zocco_alf}.  
%The ``electron shear Alf\'en wave'' corresponds to what the original shear Alfv\'en wave becomes in a pure pair plasma. 
%
Indeed, in a pure pair plasma, the dispersion relation Eq.~(\ref{alf_LD:eq}) reduces for small $k_{\perp} \lambda_D < 1$ to 
\be
\omega^2 = k_{\|}^2 \; \frac{B^2}{\mu_0 m_e n_{0e}} \; \frac{\rho_e^2}{2 \lambda_D^2} \;\; \Leftrightarrow \;\; 
\omega = k_{\|} c
\ee
As shown in Fig.~\ref{alf}, the shear Alfv\'en wave (SAW) transforms for 
$\nu_i \rightarrow 0$ into the electromagnetic wave, for which the displacement
current must be taken into account in order to address it 
properly, see \citep{Zocco_alf} for details. 
%
% WHISTLER? http://w3fusion.ph.utexas.edu/ifs/ifsreports/1159_avefiev.pdf, Eq.(8) 
% Interestingly, the dispersion relation in the pure pair plasma is quadratic in
% the wave number, which is a property of the whistler wave. Also, our result
% compares nicely with the whistler frequency, see for example Eq.~(8) in [Phys.~Plasmas, {\bf 13}, 062107 (2006)], which reads:
%\be
%\omega \approx \omega_{{\rm c}e} \frac{k_{\perp} k_{\|} c^2}{\omega_{{\rm p}e}^2} =
%\frac{1}{\sqrt{\beta_e}} (k_{\perp} \lambda_D) \, (k_{\|} c) = (k_{\|} v_A) (k_{\perp} d_e)
%\ee
%
For large $k_{\perp} \lambda_D > 1$, a whistler-type solution $\omega \sim k^2$ is obtained:
\be
2 \beta_e \zeta_e^2 = \frac{k_{\perp}^2 \rho_e^2}{2} \;\; \Leftrightarrow \;\; 
\omega = \frac{1}{2}   \; \frac{k_{\perp} \rho_e}{\sqrt{\beta_e}} \; k_{\|} v_{{\rm th}e}
\ee
This whistler-type wave can be found also in conventional plasmas and
proton-contaminated pair plasmas, as shown in Fig.~\ref{alf_LD}, where the
dispersion relation Eq.~(\ref{disp_rel}) is solved numerically.  
The transitions between the shear Alfv\'en wave, electromagnetic wave, and the
whistler can be seen clearly. 
% Note however that $\lambda_D \gtrsim \rho_e/\sqrt{\beta_e}$ implies $v_{{\rm th}e} \gtrsim c$
% \citep{Zocco_alf}. In this case, relativistic effects must be taken into
% account also in the particle dynamics (relativistic distribution functions etc).  
% For a non-relativistic regime, one has to limit the Debye length to 
% $\lambda_D / \rho_i < \sqrt{m_e/(m_i \beta_e)}$ 
% which is smaller than unity unless $\beta_e < m_e/m_i$. 
%
%
%=================================================================
%
\section{Conclusions}\label{sec:conclusions}

In this paper, we have studied the gyrokinetic stability of pair plasmas 
solving the dispersion relation (\ref{disp_rel}) analytically and
numerically. It is found that pair plasmas can support 
the gyrokinetic ITG, ETG and  universal instabilities even in slab geometry if
the proton fraction exceeds some
threshold. In practice, however, this threshold is usually quite large,
hopefully large enough to keep the proton content below this value in pair plasma
experiments \citep{pedersen2012}. These results extend the finding of
\citet{per_prl} that pair plasmas are stable to gyrokinetic modes in the
absence of magnetic curvature to the cases with small to moderate proton
contamination. We find, however, that pure pair 
plasmas can have temperature-gradient-driven instabilities, if the electron and
the positron temperature profiles differ. In reality, however, such profiles are
unlikely in steady state, since the characteristic time of energy exchange
between the species is comparable to the Maxwellisation time. 
In the electromagnetic regime, we find that the  shear Alfv\'en wave is
present in a contaminated plasma. Its frequency increases very rapidly when the
ion fraction becomes negligible.
%
%%%%%%%%%%%%%%%%%%%%%%%%%%%%%%%%%%%%%%%%%%%%%%%%%%%%%%%%%%%%%%%%%

\qquad \\
\qquad \\
{\bf Acknowledgements}
%%%%%%%%%%%%%%%%%%%%%%%%%%%%%%%%%%%%%%%%%%%%%%%%%%%%%%%%%%%%%%%%%
%
We acknowledge Thomas Sunn Pedersen and PAX/APEX experiment team for their
interest to our work. A.~M.~thanks V.~S.~Mikhailenko and V.~D.~Yegorenkov for bringing his attention to the K-mode solutions 
of kinetic dispersion relations. R.~Kleiber is acknowledged for providing a module for a numerical root finding. 
%
%=================================================================
%
%%%%%%%%%%%%%%%%%%%%%%%%%%%%%%%%%%%%%%%%%%%%%%%%%%%%%%%%%%%%%%%%%
% \appendix
% \section{}\label{appA}
% This appendix contains sample equations in the JPP style. Please refer to the
% {\LaTeX} source file for examples of how to display such equations in your
% manuscript. 

%%%%%%%%%%%%%%%%%%%%%%%%%%%%%%%%%%%%%%%%%%%%%%%%%%%%%%%%%%%%%%%%%
% susie put cite commands here, don't bother with citet etc just yet.

\bibliographystyle{jpp}
%\bibliography{elpos}

\begin{thebibliography}{12}
\expandafter\ifx\csname natexlab\endcsname\relax\def\natexlab#1{#1}\fi
\def\au#1{#1} \def\ed#1{#1} \def\yr#1{#1}\def\at#1{#1}\def\jt#1{\textit{#1}}
  \def\bt#1{#1}\def\bvol#1{\textbf{#1}} \def\vol#1{#1} \def\pg#1{#1}
  \def\publ#1{#1}\def\arxiv#1{#1}\def\org#1{#1}\def\st#1{\textit{#1}}

\bibitem[Carpentier \& Santos(1982)]{nyquist1}
{\sc \au{Carpentier, M.~P.} \& \au{Santos, A.~F.~D.}} \yr{1982}  \at{Solution
  of equations involving analytic functions.}  \jt{Journ.~Comp.~Phys.}
  \bvol{45},  \pg{210--220}.

\bibitem[Coppi {\em et~al.\/}(1967)Coppi, Rosenbluth \&
  Sagdeev]{Coppi_Rosenbluth_Sagdeev}
{\sc \au{Coppi, B.}, \au{Rosenbluth, M.~N.} \& \au{Sagdeev, R.~Z.}} \yr{1967}
  \at{Instabilities due to temperature gradients in complex magnetic field
  configurations.}  \jt{Phys.~Fluids}  \bvol{10},  \pg{582--587}.

\bibitem[Davies(1986)]{nyquist2}
{\sc \au{Davies, B.}} \yr{1986}  \at{Locating the zeros of an analytic
  function.}  \jt{Journ.~Comp.~Phys.}  \bvol{66},  \pg{36--49}.

\bibitem[Fried \& Gould(1961)]{Fried_Kmodes}
{\sc \au{Fried, B.} \& \au{Gould, R.}} \yr{1961}  \at{Longitudinal ion
  oscillations in a hot plasma}.  \jt{Phys.~Fluids}  \bvol{4}~(1),
  \pg{139--147}.

\bibitem[Helander(2014)]{per_prl}
{\sc \au{Helander, P.}} \yr{2014}  \at{Microinstability of magnetically
  confined electron-positron plasmas.}  \jt{Phys.~Rev.~Lett.}  \bvol{113},
  \pg{135003+4}.

\bibitem[Helander \& Connor(2016)]{per_jpp}
{\sc \au{Helander, P.} \& \au{Connor, J.}} \yr{2016}  \at{Gyrokinetic stability
  theory of electron-positron plasmas.}  \jt{J.~Plasma Phys.}  \bvol{82},
  \pg{9058203+13}.

\bibitem[Pedersen {\em et~al.\/}(2003)Pedersen, Boozer, Dorland, Kremer \&
  Schmitt]{pedersen2003}
{\sc \au{Pedersen, T.}, \au{Boozer, A.}, \au{Dorland, W.}, \au{Kremer, J.} \&
  \au{Schmitt, R.}} \yr{2003}  \at{Prospects for the creation of
  positron-electron plasmas in a non-neutral stellarator.}  \jt{J.~Phys B:
  At.~Mol.~Opt.~Phys.}  \bvol{36},  \pg{1029--1039}.

\bibitem[Pedersen {\em et~al.\/}(2012)Pedersen, Danielson, Hugenschmidt, Marx,
  Sarasola, Schauer, Schweikhard, Surko \& Winkler]{pedersen2012}
{\sc \au{Pedersen, T.}, \au{Danielson, J.}, \au{Hugenschmidt, C.}, \au{Marx,
  G.}, \au{Sarasola, X.}, \au{Schauer, F.}, \au{Schweikhard, L.}, \au{Surko,
  C.} \& \au{Winkler, E.}} \yr{2012}  \at{Plans for the creation and studies of
  electron–positron plasmas in a stellarator.}  \jt{New J.~Phys.}  \bvol{14},
   \pg{03510+13}.

\bibitem[Saitoh {\em et~al.\/}(2015)Saitoh, Stanja, Stenson, Hergenhahn,
  Niemann, Pedersen, Stoneking, Piochacz \& Hugenschmidt]{saitoh2015}
{\sc \au{Saitoh, H.}, \au{Stanja, J.}, \au{Stenson, E.}, \au{Hergenhahn, U.},
  \au{Niemann, H.}, \au{Pedersen, T.}, \au{Stoneking, M.}, \au{Piochacz, C.} \&
  \au{Hugenschmidt, C.}} \yr{2015}  \at{Efficient injection of an intense
  positron beam into a dipole magnetic field.}  \jt{New J.~Phys.}  \bvol{17},
  \pg{103038+9}.

\bibitem[Yegorenkov \& Stepanov(1987)]{UkrSSR}
{\sc \au{Yegorenkov, V.} \& \au{Stepanov, K.}} \at{ \yr{1987} } \jt{Dop.
  Akademii Nauk URSR, Ser. A} ~(8),  \pg{44}.

\bibitem[Yegorenkov \& Stepanov(1988)]{JETP}
{\sc \au{Yegorenkov, V.} \& \au{Stepanov, K.}} \at{ \yr{1988} } \jt{JETP}
  \bvol{94},  \pg{116}.

\bibitem[Zocco(2017)]{Zocco_alf}
{\sc \au{Zocco, A.}} \yr{2017}  \at{Slab magnetised non-relativistic low-beta
  electron-positron plasmas: collisionless heating, linear waves and
  reconnecting instabilities}.  \jt{Submitted to Journal of Plasma Physics} .

\end{thebibliography}

\end{document}